%% file: ms.tex
\documentclass[floatfix]{emulateapj} 
 
\def\aj{AJ}
\def\araa{ARA\&A}
\def\apj{ApJ}
\def\apjl{ApJL}
\def\apjs{ApJS}
\def\ao{Appl.~Opt.}
\def\apss{Ap\&SS}
\def\aap{A\&A}
\def\aaps{A\&AS}
\def\mnras{MNRAS}
\def\pasp{PASP}
\def\ssr{Space~Sci.~Rev.}
 
\DeclareMathAlphabet{\mathsc}{OT1}{cmr}{m}{sc} 
\def\testbx{bx}%
\DeclareRobustCommand{\ion}[2]{%
\relax\ifmmode 
\ifx\testbx\f@series 
{\mathbf{#1\,\mathsc{#2}}}\else 
{\mathrm{#1\,\mathsc{#2}}}\fi 
\else\textup{#1\,{\mdseries\textsc{#2}}}%
\fi} 
\newcommand{\ha} {\mbox{H$\alpha$}} 
\newcommand{\hb} {\mbox{H$\beta$}}

\newcommand{\Feii} {\ion{Fe}{ii}} 
 
\newcommand{\Caiia} {[\ion{Ca}{ii}]} 
\newcommand{\Cai} {\ion{Ca}{i}} 
\newcommand{\Caii} {\ion{Ca}{ii}} 
 
\newcommand{\Bai} {\ion{Ba}{i}} 
\newcommand{\Baii} {\ion{Ba}{ii}} 
\newcommand{\Nai} {\ion{Na}{i}} 
 
\newcommand{\Mgii} {\ion{Mg}{ii}} 
 
\newcommand{\Hii} {\ion{H}{ii}} 
\newcommand{\Scii} {\ion{Sc}{ii}} 
 
\newcommand{\Tiii} {\ion{Ti}{ii}} 
\newcommand{\Hei} {\ion{He}{i}}

\newcommand{\Oia} {[\ion{O}{i}]} 
\newcommand{\Oi} {\ion{O}{i}}

\def\sn{SN 2008in}

\newcommand{\ebv}{\mbox{$E(B-V)$}}

\newcommand{\msun}{\mbox{$M_{\odot}$}}

\newcommand{\rsun}{\mbox{$R_{\odot}$}} 
\newcommand{\kms}{\mbox{$\rm{\,km\,s^{-1}}$}}

\newcommand{\nickel}{\mbox{$^{56}$Ni}}

\newcommand{\mum}{\mbox{$\mu{\rm m}$}}

\shorttitle{SN 2008in IN M 61} 
\shortauthors{ROY ET AL.} 
 
\begin{document} 
 
 
\title{\sn\ $-$ Bridging the gap between normal and faint supernovae of type IIP} 
 
 
\author{ 
Rupak Roy$^\star$\altaffilmark{1},  
Brijesh Kumar\altaffilmark{1},  
Stefano Benetti\altaffilmark{2},  
Andrea Pastorello\altaffilmark{3},  
Fang Yuan\altaffilmark{4,5},  
Peter J. Brown\altaffilmark{6},  
Stefan Immler\altaffilmark{7,8},  
Timur A. Fatkhullin\altaffilmark{9},  
Alexander S. Moskvitin\altaffilmark{9},  
Justyn Maund\altaffilmark{10},  
Carl W. Akerlof\altaffilmark{4},  
J. Craig Wheeler\altaffilmark{11},  
Vladimir V. Sokolov\altaffilmark{9},  
Rorbert M. Quimby\altaffilmark{12},  
Filomena Bufano\altaffilmark{2},  
Brajesh Kumar\altaffilmark{1,13},  
Kuntal Misra\altaffilmark{14,15},  
S. B. Pandey\altaffilmark{1,4},  
Nancy Elias-Rosa\altaffilmark{16},  
Peter W. A. Roming\altaffilmark{17} and  
Ram Sagar\altaffilmark{1} 
} 
 
\email{$^\star$roy@aries.res.in, rupakroy1980@gmail.com}  
 
\altaffiltext{1}{Aryabhatta Research Institute of Observational Sciences (ARIES), Manora 
    Peak, Nainital, 263 129, India} 
 
\altaffiltext{2}{Istituto Nazionale di Astrofisica, Observatorio Astronomico di Padova, 
    Italy} 
 
\altaffiltext{3}{Astrophysics Research Centre, School of Mathematics and Physics, 
    Queen’s University Belfast, Belfast BT7 1NN, UK} 
 
\altaffiltext{4}{Randall Laboratory of Physics, University of Michigan, 450 Church Street, 
   Ann Arbor, MI 48109-1040, USA} 
 
\altaffiltext{5}{Research School of Astronomy and Astrophysics, The Australian National 
   University, Cotter Road, Weston Creek, ACT 2611,Australia} 
 
\altaffiltext{6}{Department of Physics and Astronomy, University of Utah, Salt Lake City, UT 84112, USA} 
 
\altaffiltext{7}{NASA/Goddard Space Flight Center, Astrophysics Science Division, Code 662, 
    Greenbelt, MD 20771, USA} 
 
\altaffiltext{8}{Department of Astronomy, University of Maryland, College Park, MD 20742, USA} 
 
\altaffiltext{9}{Special Astrophysical Observatory, Nizhnij Arkhyz, Karachaevo-Cherkesia, 
    369167 Russia} 
 
\altaffiltext{10}{Dark Cosmology Centre, Niels Bohr Institute, University of Copenhagen, 
    Juliane Maries Vej 30, 2100 Copenhagen, Denmark} 
 
\altaffiltext{11}{Department of Astronomy, University of Texas, Austin, TX 78712–0259, USA} 
 
\altaffiltext{12}{Cahill Center for Astrophysics, California Institute of Technology, 
     Pasadena, CA 91125, USA} 
 
\altaffiltext{13}{Institut d'Astrophysique et de G\'{e}ophysique, 
    Universit\'{e} de Li\`{e}ge, All\'{e}e du 6 Ao\^{u}t 17, B\^{a}t B5c, 
    4000 Li\`{e}ge, Belgium} 
 
\altaffiltext{14}{Space Telescope Science Institute, 3700 San Martin Drive, Baltimore, MD 
    21218, USA} 
 
\altaffiltext{15}{Inter University Center for Astronomy and Astrophysics, Post Bag 4, 
    Ganeshkhind, Pune, 411 007, India} 
 
\altaffiltext{16}{Institut de Ci$\check{c}$ncies de I'Espai (IEEC-CSIC),
    Campus UAB, 08193 Bellaterra, Spain} 
 
\altaffiltext{17}{Space Science and Engineering Division, 6220 Culebra Rd, San Antonio, TX 78238-5166, USA}

\begin{abstract}
 We present optical photometric and low-resolution spectroscopic observations
 of the Type II plateau supernova (SN) 2008in, which occurred in the outskirts of the
 nearly face-on spiral galaxy M 61. Photometric data in the X-rays, ultraviolet and
 near-infrared bands have been used to characterize this event. The SN field was imaged
 with the ROTSE-IIIb optical telescope about seven days before the explosion. This allowed us to
 constrain the epoch of the shock breakout to JD = 2454825.6. The duration of the plateau phase, as
 derived from the photometric monitoring, was $\sim$ 98 days.
 The spectra of \sn\ show a striking resemblance to those of the archetypal low-luminosity IIP SNe 1997D and 1999br.
 A comparison of ejecta kinematics of \sn\ with the hydrodynamical simulations of Type IIP SNe by \citet{dessart10}
 indicates that it is a less energetic event ($\sim 5\times10^{50}$ erg).
 However, the light curve indicates that the production of radioactive $^{56}$Ni is significantly
 higher than that in the low-luminosity SNe. Adopting an interstellar absorption
 along the SN direction of $A_V \sim$ 0.3 mag and a distance of 13.2 Mpc, we estimated a synthesized
 \nickel\ mass of $\sim0.015 M_{\sun}$. Employing semi-analytical formulae \citep{litvinova85}, we derived a
 pre-SN radius of $\sim 126$\rsun\,, an explosion energy of $\sim 5.4\times10^{50}$ erg and a total ejected mass of
 $\sim 16.7$\msun\,. The latter indicates that the zero age main-sequence mass of the
 progenitor did not exceed 20\msun\,. Considering the above properties of \sn\,
 and its occurrence in a region of sub-solar metallicity ([O/H] $\sim$ 8.44 dex),
 it is unlikely that fall-back of the ejecta onto a newly formed black hole occurred in SN 2008in. We
 therefore favor a low-energy explosion scenario of a relatively compact, moderate-mass progenitor star
 that generates a neutron star.
 \end{abstract}

 
 



\section{introduction} \label{intro} 

 Core-collapse Type II supernovae (SNe) mark the violent death of stars with main-sequence 
 masses greater than 8 \msun\ and as is indicated by the presence of hydrogen lines in their 
 optical spectra, they originate from a progenitor star with a significant amount of hydrogen 
 still intact \citep{eldridge04}. Of special interest are the Type II plateau SNe which are characterized 
 by a `plateau' in their optical light curve and are more common, constituting about 75\% of all Type II SNe
 \citep{smith10}. The IIP SNe show a wide range of plateau luminosities, plateau durations, expansion velocities
 and nickel masses \citep{hamuy03} and these observational properties are connected with the explosion mechanism
 as well as the physical properties of the progenitor star such as ejected mass, explosion energy and 
 pre-SN radius \citep{nadyozhin03, smartt09}. The IIP SNe are thought to result from progenitor masses in the 
 range $8-25$ \msun\, \citep{heger03} with an extended hydrogen envelope necessary to maintain the plateau phase. 
 A detailed study of optical light curves and spectra of only a few nearby IIP SNe has been done so far 
 and there exists a discrepancy in estimating the mass of their progenitors, e.g, for the three well studied 
 events (namely 1999em, 2005cs and 2004et), the determination of progenitor mass
 from the hydrodynamical modeling of their light curve is found to be higher than that estimated from 
 pre-SN imaging \citep{utrobin10, bersten11}.

 Recently, a number of `low luminosity' Type IIP events have been discovered viz.  SNe 1999br, 1999eu, 
 1994N, 2001dc, 2005cs \citep{pastorello04, pastorello09}, 2008bk \citep{vandyk11}
 and 2009md \citep{fraser10}. These events have  explosion energy ($\sim 10^{50}$ erg) and 
 ejected $^{56}$Ni-mass ($2-7\times10^{-3}$ \msun\,), both lower by a factor of 
 10 than normal, and low expansion velocity $\sim$1000 \kms (\citealt{pastorello09} and references therein). 
 The low-luminosity IIP SNe are debated because of the unknown nature of their progenitors. The 
 first reported faint SN was SN 1997D \citep{turatto98, benetti01}, and the observed properties of 
 its light curves and spectra  were explained in terms of significant fallback of ejected material 
 on a newly formed black hole (BH), created through the core collapse of a massive progenitor 
 ($M \ga 20M_\odot$, \citealt{zampieri98,zampieri03}). Alternatively, SN 1997D was interpreted 
 as the explosion of a less massive progenitor ($8-12M_\odot$, \citealt{chugai00}), 
 close in mass to the lower limit for stars that can undergo core-collapse. 
 \citet{heger03} suggested that low-luminosity Type IIP events are electron capture 
 SNe produced by low-mass progenitors giving rise to ONeMg cores. This is further 
 supported through investigations of pre-explosion images \citep{maund05a,maund05b},
 though \citet{eldridge07} have ruled out the possibility of such a mechanism 
 for the low-luminosity SN 2005cs. According to the formalism of \citet{heger03} and 
 \citet{eldridge04}, no star having initial mass less than 22\msun\ can form a BH,
 which can quench the ejected material and produce a low luminosity SN.
 Stars with masses above 25$M_\odot$, formed in metal-poor or slightly 
 sub-solar metallicity regions can produce low-luminosity, BH-forming Type IIP SNe. 
 Type IIL/b events can be produced through this process from stars having masses 
 $\textgreater~ 25$\msun\ and generated in regions with solar (or super-solar)
 metallicity. So, the metallicity information at the SN location and the estimation 
 of the initial mass are essential to constrain the triggering mechanisms of these 
 explosions.  
  
 SN 2008in was discovered in the nearby galaxy M 61 (NGC 4303). The first 
 unfiltered CCD images of \sn\ were taken by Koichi Itagaki on 2008 December 26.79 
 (all times in UT hereafter) and 27.69 at a magnitude of 14.9. Independent observations of this 
 event by K. Kadota showed the transient at an unfiltered mag of 15.1, In addition 
 W. Wells recorded the SN on 2008 December 28.46  at $V$ and $R$ band magnitudes of 14.3 
 and 13.2, respectively \citep{nakano08}. Low and mid-resolution spectroscopic 
 observations indicate an early discovery for SN 2008in (within 1$-$2 weeks after 
 core-collapse). The spectra showed highly blueshifted H$_\alpha$ and H$_\beta$ 
 absorptions (by $\sim$ 9000~\kms\,) with weaker emission components 
 \citep{chakraborti08,foley08,stritzinger08}. The presence of prominent 
 P-Cygni profiles of Balmer lines leads to its classification as a Type II SN. 

 The broadband light curve and the initial spectral evolution of SN 2008in were 
 similar to those of normal Type IIP SNe. However, from mid-plateau, the SN 
 started to show a few spectral features (like \ha\,) which are similar to
 under-luminous events. \sn\ was also observed in the radio with the Very Large 
 Array (VLA) on 2008 December 31.40 UT in two frequency bands at 8.4601 and 22.4601 
 GHz \citep{Stockdale08}. Observation for the second epoch was further reported 
 by \citet{Stockdale09} on 2009 January 27. Interacting Type II SNe, like Type IIn 
 events and a few Type IIP events (e.g. SN 2004et) are supposed to be strong sources 
 of radio emission (for review see \citealt{weiler02}). However,
 both VLA observations produced null results for this proximate event.
 
 In this paper we present optical and near-infrared photometric and 
 optical spectroscopic observations of \sn. The photometric data cover 
 a time span of about 410 days since the discovery. The {\it Swift}/XRT \citep{burrows05} and 
 {\it Swift}/UVOT \citep{roming05} data covering a time span of 60 days are also presented.  
 In Sections~\ref{res:broadband} and~\ref{res:spec} we study the  
 the photometric and spectroscopic evolutions respectively. 
 In Sections~\ref{res:DistExt} and~\ref{res:ColBol} we cover the estimates of 
 distance, reddening, intrinsic color and bolometric fluxes. The main physical parameters 
 of the explosion and the mass of the progenitor are derived in Section~\ref{res:parameter}
 while a comparison of its properties with other SNe is given in Section~\ref{res:comp}.
 The conclusion of the paper is given in the last section. 
 The epoch of explosion JD = 2454825.6 (\S\ref{res:broadband}) is considered throughout the paper 
 and the times of pre-/post-explosion are referred with $-/+$ signs respectively.


\section{observation and data reduction} \label{obs} 
 
\subsection{Photometric Observation} \label{obs:phot} 

 The prompt follow-up of the event was carried out by the ground-based 
 ROTSE-IIIb telescope\footnote{The Robotic Optical Transient Search 
 Experiment (ROTSE-III) is a set of four 45 cm fully robotic optical telescopes, 
 installed at Siding Spring Observatory, Australia (ROTSE-IIIa), McDonald Observatory, 
 Texas (ROTSE-IIIb), H.E.S.S. site, Namibia (ROTSE-IIIc) and TUBITAK National 
 Observatory, Turkey (ROTSE-IIId; \citealt{akmrs03,yuan_thesis}).} having sensitivity in 
 the wavelength region from 0.35 to 1.0 \mum\ with a peak around 0.6 \mum\
 \citep{quimby07}. The SN was first detected in the 
 ROTSE-IIIb images on 2008 December 24.45 and it was monitored at 58 phases 
 until +115d. The initial detections of the SN evaded the automated pipeline 
 identification due to poor image quality and low signal-to-noise ratio (S/N). 
 The data reduction was performed afresh on all the available ROTSE data. 
 In order to remove the contamination of the true SN-flux from the underlying 
 galaxy, a pre-SN galaxy template was constructed from images taken in early 2008 and 
 each SN frame was reduced using the galaxy-template subtraction scheme developed by
 \citet{alard00}. The point spread function (PSF) photometry was performed at the SN location 
 in the galaxy-template subtracted images. The unfiltered instrumental magnitudes 
 were calibrated using the USNO B1.0 $R$-band magnitudes of about 15 isolated stars. 
 The light curve thus produced was found to be 0.15 mag off from the Cousins $R$ 
 band light curve produced by multi-band observation carried out at ARIES 
 (described below) and the ROTSE magnitudes were scaled accordingly.
  
\input{uvot.tex} 

 The SN 2008in was also monitored with the Ultraviolet Optical Telescope (UVOT) on 
 board {\it Swift} from +5d to +60d. The UVOT filters $uvw2, uvm2, uvw1, u, b$, and $v$
 have their effective wavelength at 2030, 2231, 2634, 3501, 4329, and 5402 \AA,
 respectively (Poole et al. 2008). The UVOT data reduction was performed following
 the prescriptions of \citet{brown09a}. A 5\arcsec\ aperture is used to measure the 
 counts for the coincidence loss correction whereas a 3\arcsec\ aperture was used for 
 the photometry. For the filters $uvw2, uvm2$, and $uvw1$, the last epoch (obtained about 60 days 
 after explosion when the UV flux was very weak) data were used to subtract the galaxy 
 light, while in the optical the SN remains much brighter than the underlying light 
 so contamination was negligible. An aperture correction (based on an average PSF in
 {\it Swift} CALDB) as well as zeropoints from \citet{poole08} was applied to put 
 the magnitudes on the UVOT photometric system. The UVOT 
 magnitudes are listed in Table~\ref{tab:uvotsn}. 

\begin{figure*} 
\includegraphics[width=18cm]{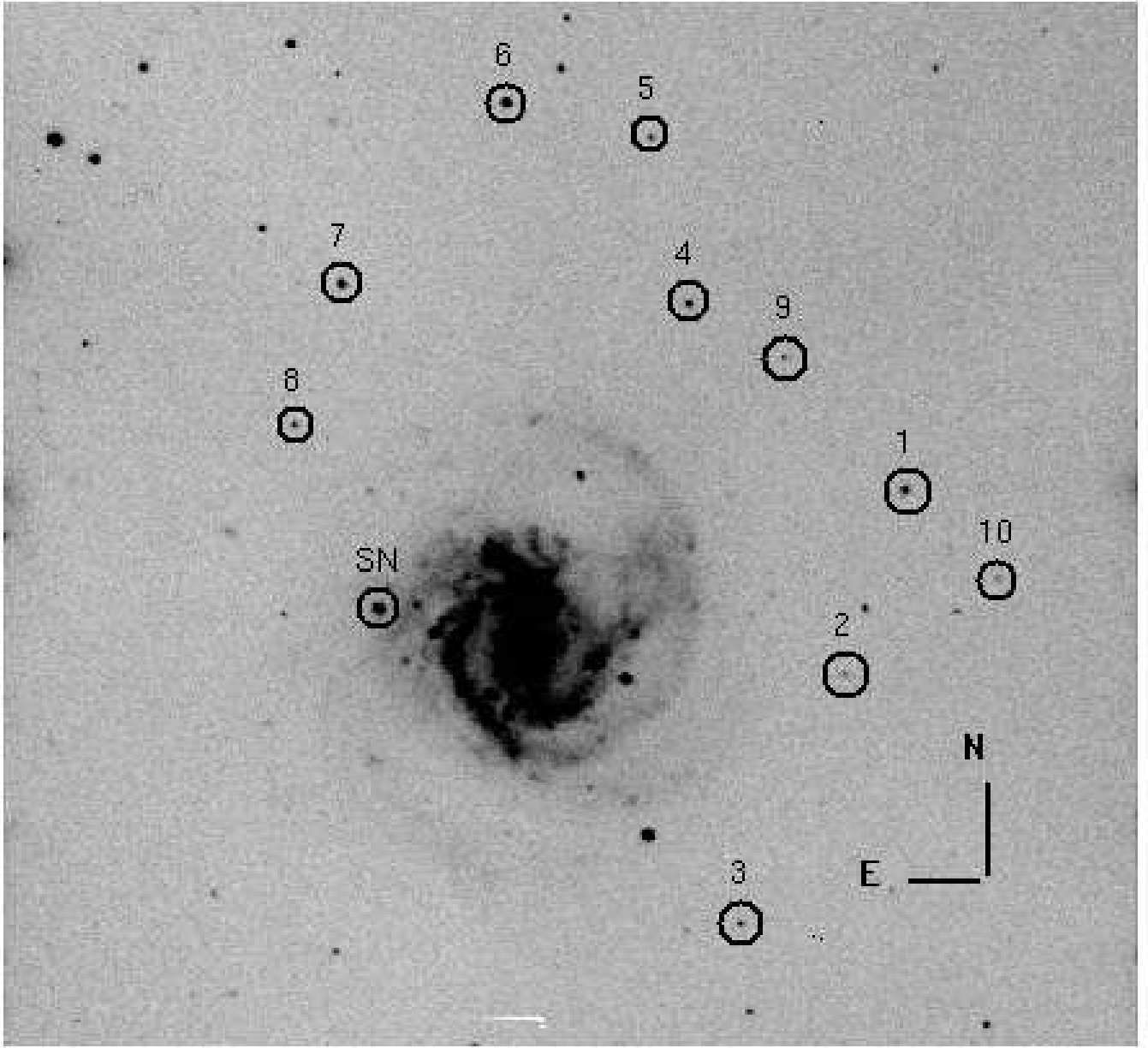}%
\caption{SN 2008in in M 61. A 600 s $V$-band image taken at phase +32d from the 1-m ST ARIES, India and covering 
         an area of about 10\arcmin $\times$10\arcmin\, is shown. The location of \sn\ as well as the local standard 
         stars are marked with circles. North is up and east is to the left.} 
\label{fig:snid} 
\end{figure*}

\begin{figure*} 
\centering
\includegraphics[width=5cm]{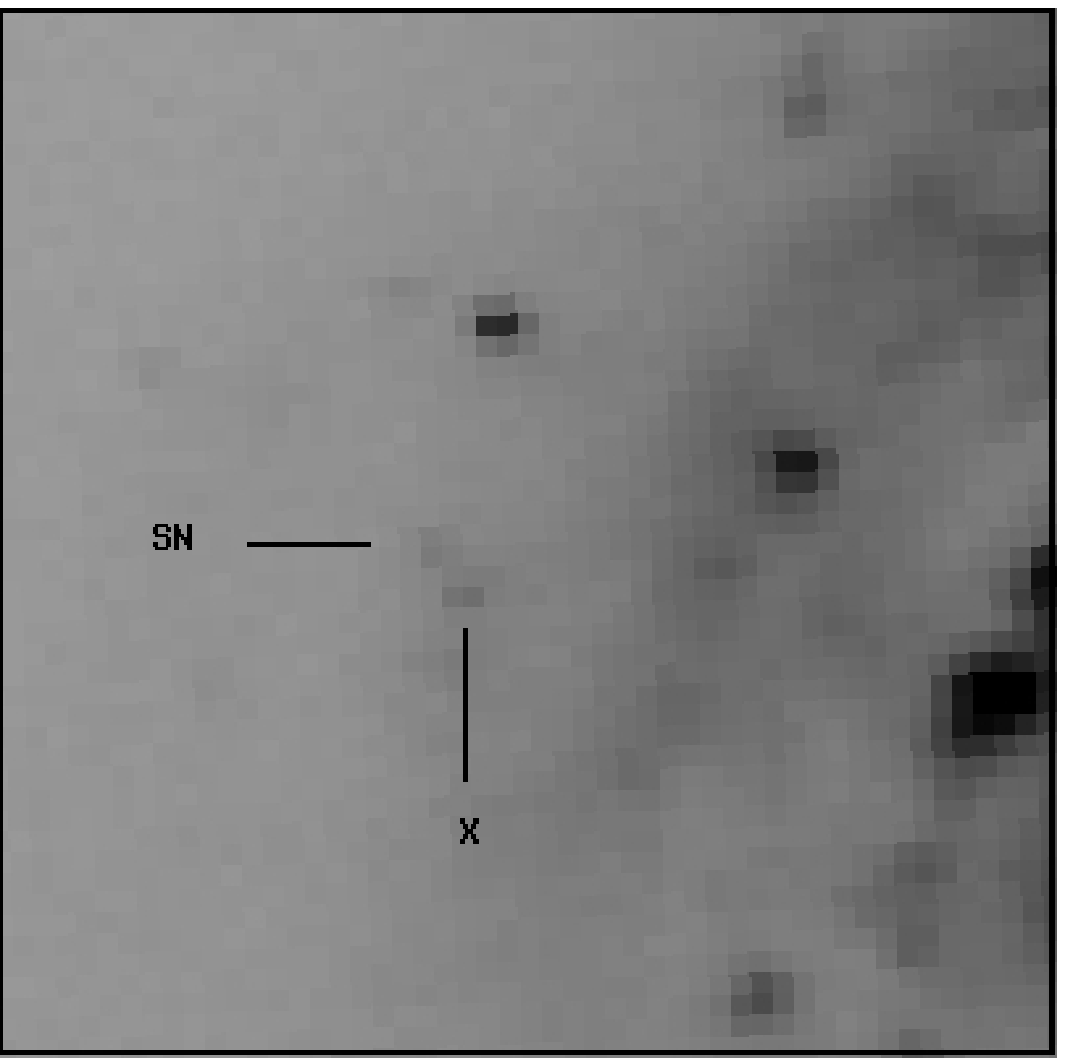}%
\includegraphics[width=5cm]{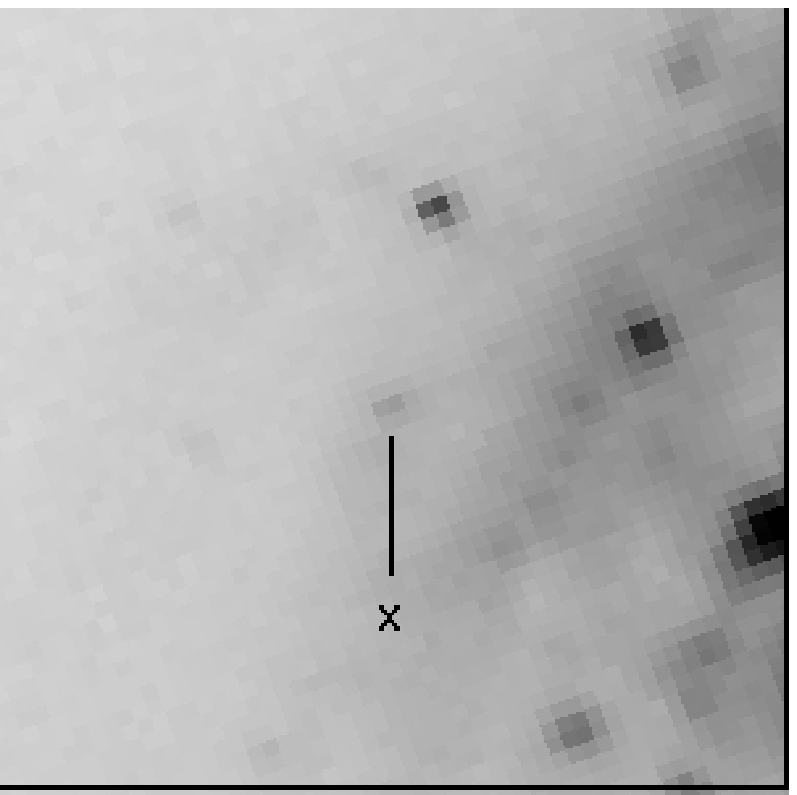}%
\includegraphics[width=5cm]{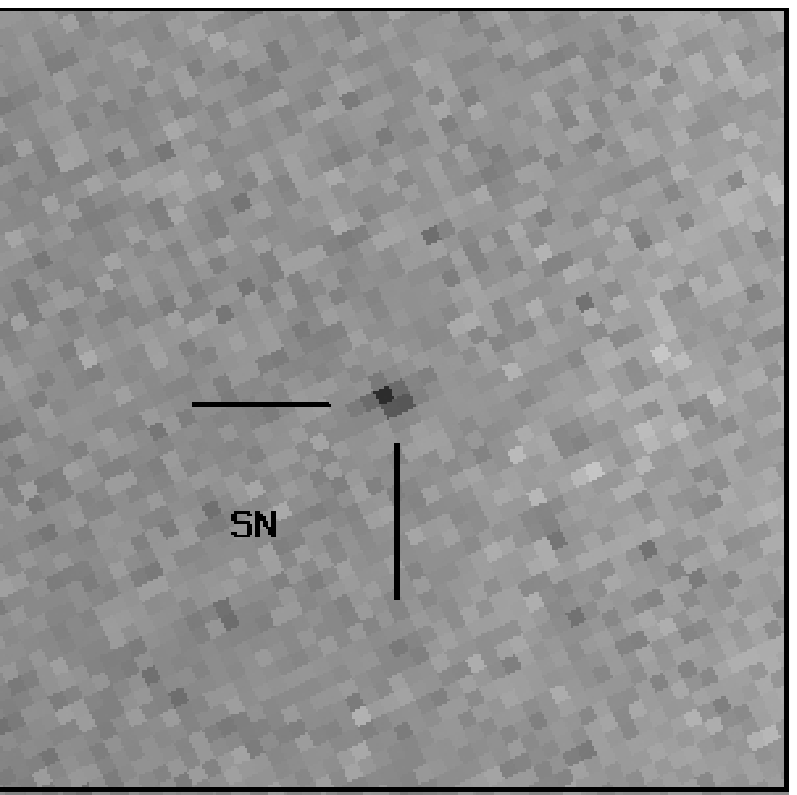}%
\caption{Measurement procedure of the true SN flux in the nebular phase. Each image shows $76\arcsec\times76\arcsec$ 
 around SN location. North is up and east is to the left. The leftmost panel shows $V$-band image observed 
 on 2010 January 20 with a SN in it. The middle panel shows the template image taken on 2011 January 04, 
 with no SN in it. A small flux enhancement marked with a cross ($\times$) symbol is probably due to a 
 star forming region. The rightmost panel shows the subtracted image, where only SN is present.} 
\label{fig:subimg} 
\end{figure*}

 From +5d to +416d, the follow-up of SN 2008in in optical broadband Johnson $UBV$ 
 and Cousins $RI$ was performed with the 1-m Sampurnanand Telescope (ST) at ARIES, 
 Nainital\footnote{ A 2048 $\times$ 2048 CCD camera mounted at the f/13 Cassegrain 
 focus of the telescope, has a square pixel of 24 micron on a side, and with a plate 
 scale of 0.38 arcsec per pixel, it covers a square area of 13$\arcmin$ on a side 
 in the sky. The gain and readout noise of the CCD camera are 10 electrons per 
 analog-to-digital unit and 5.3 electrons, respectively. To improve the 
 S/N ratio and optimize the sampling, the observations were carried out in a binned mode of 
 2$\times$2 pixel.}. 
 An identification chart showing the field of the galaxy 
 M61 along with the locations of \sn\ as well as the local standards is 
 presented in Figure~\ref{fig:snid}. The photometry is performed  
 using standard tasks of IRAF \footnote{IRAF 
 stands for Image Reduction and Analysis Facility distributed by the National Optical 
 Astronomy Observatory which is operated by the Association of Universities for 
 research in Astronomy, Inc. under co-operative agreement with the National Science 
 Foundation} and {\it DAOPHOT} \footnote{ {\it DAOPHOT} stands for Dominion 
 Astrophysical Observatory Photometry \citep{stetson87}} as described in the paper
 by \citet{roy11}. Bias subtraction and flat fielding 
 were performed on the raw frames and the cosmic removal was done using LACOSMIC routine
 \citep{dokkum01}. As the SN lies in the outskirts of the galaxy on a relatively faint and
 smooth background, the photometry at the initial 
 phases (mostly during the plateau phase when the SN is bright) is estimated using the profile 
 fitting method. During nebular phases, when the SN becomes faint, the true SN flux is estimated using 
 the galaxy template subtraction method following the procedures of \citet{roy11}. 
 As a galaxy template, we used post-explosion (+600d) images observed on 2011 January 04 under good 
 seeing conditions. Figure~\ref{fig:subimg} shows an example of this procedure applied to
 a late-time $V$-band image of \sn. In the template image of 2011 January 04, we can see a 
 clear flux enhancement located approximately 5\arcsec away (a linear distance of $\sim$ 343pc) 
 from the SN position. This knot is also present in a deep image from the Sloan Digital Sky
 Survey (SDSS) and it is identified as a \Hii\ 
 region in the NED catalog. The progenitor of \sn\ may have a possible association with this
 star-forming region. The field of \sn\ was calibrated using \citet{landolt09} standard stars of the 
 field SA98 observed on the same night as the SN. A sample of 10 bright and isolated non-variable stars 
 in the field of \sn\ was used as local standards to derive the zero points for the SN at each epoch. 
 The location and magnitudes of these local standards are listed in Table \ref{tab:photstar}. The entire 
 time span of photometric observation is about 410 days and in Table \ref{tab:photsn}, we report the 
 $UBVRI$ photometry of the event. 
  
 From +4d to +116d, the \sn\ was also observed in $JHK$ near-infrared (NIR) bands with the 0.6-m REM/REMIR at 
 La Silla \citep{zerbi04}. The object was clearly visible in the $J$ and $H$ passbands, while it remained undetected 
 in $K$. The instrumental magnitudes were standardized using the 2MASS standards available in the field of \sn. The 
 calibrated $JH$ magnitudes are listed in Table \ref{tab:remsn}.

\subsection{X-ray Observations} \label{obs:xrt} 

 The {\it Swift} XRT observations were obtained simultaneously 
 with the UVOT observations. To search for X-ray emission from \sn\,, we extracted X-ray 
 counts from a circular region with a 10~pixel radius ($23\farcs7$, corresponding to the XRT 
 on-axis 90\% encircled energy radius) centered on the optical position of the SN. The background 
 was extracted locally from a source-free region of $40\arcsec$ radius to account for the detector,
 sky background, and the diffuse emission from the host galaxy.  
 
 No X-ray source is detected in the merged 27.1~ks XRT data obtained in photon-counting mode. The 
 $3\sigma$ upper limit to the XRT net count rate is $7.2 \times 10^{-4}~{\rm cts~s}^{-1}$, corresponding 
 to an unabsorbed (0.2--10~keV band) X-ray flux of 
 $f_{0.2-10} < 3.4 \times 10^{-14}~{\rm erg~cm}^{-2}~{\rm s}^{-1}$ and a luminosity of 
 $L_{0.2-10} < 7.0 \times 10^{38}~{\rm erg~s}^{-1}$ for an adopted thermal plasma spectrum with a 
 temperature of $kT = 10~{\rm keV}$ (see \citealt{fransson96} and references therein), 
 a Galactic foreground column density of $N_{\rm H} = 1.67 \times 10^{20}~{\rm cm}^{-2}$ \citep{dickey90} 
 and a distance of 13.19~Mpc (\S\ref{res:DistExt}).


\input{sec_star.tex} 

\input{phot_sn.tex} 
 
\input{rem.tex} 
 
\input{speclog.tex} 

\subsection{Optical Spectroscopic Observation} \label{obs:spec} 

 Long-slit low-resolution spectra ($\sim$ 6 to 14 \AA\,) in 
 the optical range (0.33 $-$ 1.0 \mum) were collected at eleven phases from +7d to +143d, including 
 five phases from the 2-m IGO, three phases from the 9.2-m HET, two phases from the 6-m BTA and 
 one epoch from the 3.6-m NTT \footnote{The full names of the telescope are provided in Table~\ref{tab:speclog}}.
 A journal of spectroscopic observations is given in Table~\ref{tab:speclog}. 
 The spectroscopic data acquired from IGO, NTT and HET were reduced under the IRAF environment. 
 Bias and flat-fielding were performed on all the frames. Cosmic-ray rejection was 
 done using the Laplacian kernel detection method \citep{dokkum01}. All the data obtained 
 from the BTA were reduced using programs in the IDL software environment. 
 
 The instrumental FWHM resolution of 2-m IGO spectra as measured from the \Oia\,$\lambda$5577\  
 emission skyline was found to lie between 6 and 10\AA\ ($\sim$ 322 - 510 \kms).  
 Flux calibration was done using standard spectrophotometric fluxes from \citet{hamuy94},  
 assuming a mean extinction for the site. For HET, BTA and NTT, the resolution 
 near 6000\AA\ is about 10\AA, 14\AA\,, and 12\AA\, respectively.


\section{Light Curve Evolution} \label{res:broadband}  

 According to theoretical interpretation, the entire broadband evolution of Type II SNe can be 
 segmented into three phases: the rising phase, the plateau phase and  the nebular phase. 


\begin{figure*} 
\centering 
\includegraphics[width=13cm]{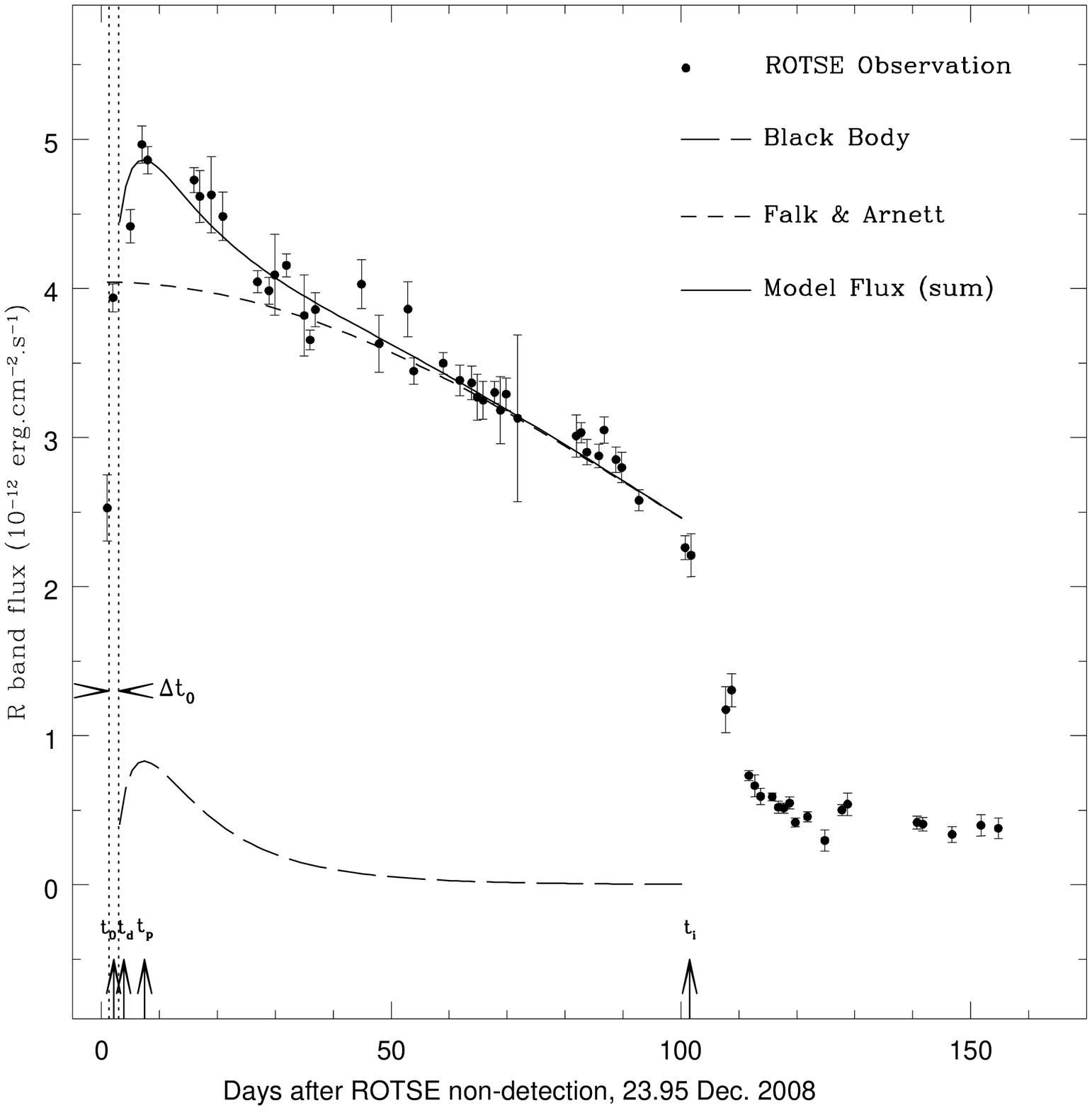}%
\caption{The ROTSE $R$-band light curve of \sn. Reference time for x-axis is the epoch of ROTSE non detection on
 2008 December 23.95. The evolution of post shock breakout flux has been modeled by using the simple analytical 
 expressions given in \citet{cowen10} and \citet{arnett80}. The best fit time for shock breakout is marked
 with $t_0$ while the $\Delta t_0$ shows the duration of uncertainty. The $t_d$ is the time of discovery 
 reported by Nakano et al. 2008. The $t_p$ is the time when $R$ light shows the peak. The $t_i$ is the time of 
 inflection obtained using the procedure of \citet{elmhamdi03b}.} 
\label{fig:rotse} 
\end{figure*}
 
\subsection{The Rising Phase} \label{res:rise} 

 The rising phase of the light curve is associated with the shock breakout phenomenon
 having been theoretically predicted for an SN explosion, in which, the 
 radiation-dominated shock wave, generated through the 
 reversal of iron core-collapse, starts to propagate outward through the onion shell-like 
 layers of the progenitor; when the shock reaches regions with an optical depth of a few tens to unity, 
 the radiation behind the shock escapes the outer surface giving rise  
 to a hot (T $>10^5$ K) fireball which emits quasi-black body radiation in UV and 
 soft-Xrays. This phenomenon is called shock breakout and depending on envelope mass, 
 density structure and wind properties of the progenitor, the breakout light curve may last from
 a few hours to few a days (\citealt{grassberg71, chevalier76, falkarnett77}). Due to the short timescale, the
 detection of shock breakout is rare, and only recently were, the UV light curves of the entire shock breakout phase
 lasting several hours was reported for two Type IIP SNe $-$ SNLS-04D2dc and SNLS-06D1jd (\citealt{gezari08, schawinski08}).
 Observation of the earliest UV and optical light curves of IIP SNe is crucial to model the shock breakout light curves
 and constrain the properties of SN progenitors \citep{tominaga09}. The earliest SNe IIP optical light curves 
 have been studied in the past for SN 2005cs by \citet{pastorello09} and for SN 2006bp by \citet{quimby07} and 
 the optical rising phase is rarely observed. 


\begin{figure*}
\includegraphics[width=15cm]{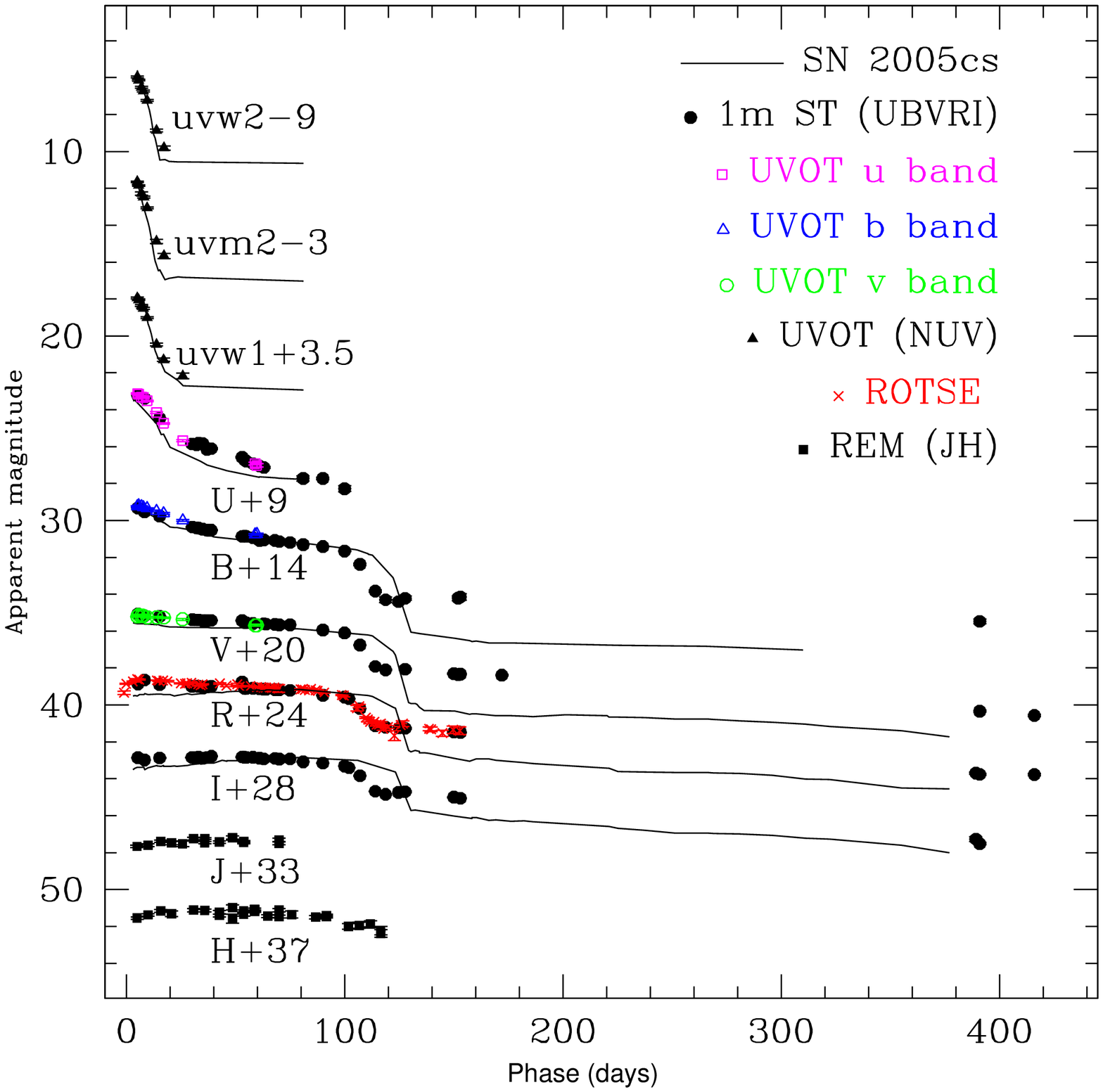}%
\caption{Light curves of \sn\, in the UV, optical and NIR bands. The light curves of archetypal low-luminosity SN 2005cs 
  are overplotted for comparison \citep[data are taken from][]{brown09b,pastorello09}. The light curves of SN 2005cs are scaled 
  in magnitude to match the light curves of SN 2008in.} 
\label{fig:lightcur} 
\end{figure*} 

 We model the early optical (ROTSE-IIIb $R$-band) light curves of \sn\, using a simple model and put an observational 
 constraint on the epoch and duration of shock breakout. Following the formulation 
 of \citet{waxman07}, it can be shown that just after the shock breakout, the intensity of the SN light
 at a fixed wavelength is proportional to the  intensity of blackbody radiation at that 
 wavelength (see Equation 1 of \citealt{cowen10}), while for the plateau phase \citet{arnett80} derived 
 an analytical expression (Equation 39 of that paper) and hence by combining these two equations one can 
 approximate the overall intensity profile of a Type II SN during the rising phase through the plateau 
 with the following expression : 

  \begin{math} 
  I_{SN}(t) = \frac{A}{exp[B\times(t-t_0)^{0.5}]-1}\times(t-t_0)^{1.6} 
  \end{math}    
  \begin{equation} 
   + C\times exp[-\{(t/D) + (t/E)^2\}] 
  \end{equation}    
 
 Here, the first term represents the phase associated with the shock breakout while the second term represents 
 the plateau phase. 
 Here `$t$' is the time measured in days since 2008 December 23.95 (JD = 2454823.5), the epoch when ROTSE had a 
 non-detection of the SN with limiting magnitude 16.16 in the $R$ band. The time of shock breakout, `$t_0$' 
 along with the constants $A, B, C, D$ and $E$ are free parameters of the fit. The values of these constants 
 depend on the nature of the progenitor and the kind of explosion. The above model is fitted to the 
 ROTSE data using the $\chi$ minimization technique \citep{press92} and the fit is shown in Figure~\ref{fig:rotse}.
 We obtained a value of the shock breakout time as  $t_0 = 2.13\pm0.83$ day. This value of $t_0$ is 
 consistent with the first data point observed from ROTSE. We shall adopt $t_0$, corresponding to 
 JD = 2454825.6 as the explosion epoch for all the phases of \sn. It is noted that at $V$ and $R$ bands,
 the SN was detected by \citet{nakano08}, just 2.4 d after the shock breakout. We note, however, that 
 depending on the extent of the envelope the 
 true core reversal marking the SN explosion would have occurred a few hours earlier
 (e.g., SN 1987A) or a few days earlier for a red supergiant envelope $\sim 50000\rsun$ \citep{quimby07}.

 In Figure~\ref{fig:rotse}, `$t_{p}$' (+5.3 d) corresponds to the peak of the ROTSE $R$-band light curve. 
 A similar peak in the UV light curves of the post-shock breakout phase of the Type IIP supernova 
 SNLS-04D2dc lasting several days was identified as a secondary peak and this 
 peak was explained as the shift of the spectral energy distribution toward longer wavelengths 
 due to the rapid fall in temperature during free adiabatic expansion of optically thick plasma lead 
 by a collisionless shock (\citealt{schawinski08, tominaga09}). For \sn\, we identify the peak in 
 the ROTSE $R$-band curve as a secondary peak and also speculate that for a few other type IIP SNe such 
 as 1999em, 1999gi and 2005cs, where very tiny peaks before the plateau light curve were observed, such
 features are basically the footprints of the secondary peaks which could be a consequence of shock breakout.

\subsection{The Plateau and Nebular Phase} \label{res:PlatNeb} 

 The entire UV, optical and NIR light curves of \sn\ are shown in Figure \ref{fig:lightcur} and for
 comparison, the light curves of the archetypal low-luminosity SN 2005cs \citep{brown09b, pastorello09} 
 are also plotted by scaling them in magnitude to match it with the observed plateau of \sn. In the
 early plateau phase, a rapid drop in the UV flux and a slowly declining or constant optical and NIR
 flux are clearly apparent. The $uvm2$  
 magnitude declined from about 14.7 to 18.7 mag within 12 days. The decline rates of 
 the \sn\, flux in UVOT, $U$, $B$, and $V$ bands are almost identical with that of SN 2005cs. 
 Starting from the $uvw2$ until the $R$ band, the measured decline rates from shock breakout to the 
 plateau phase are approximately 0.32, 0.34, 0.25, 0.13, 0.04, 0.01 and 0.01 mag d$^{-1}$. The $IJH$ 
 light curves are almost flat even at early phases of the plateau. For this SN, the plateau is well 
 sampled in the $BVRI$ bands, so we can accurately determine the plateau duration. As discussed in 
 \S\ref{res:rise}, the secondary peak ($t_{p} \sim +5.3$ d) in the prompt light curve is generated by the 
 gradually 
 cooling shock heated SN atmosphere and after that the plateau mechanism starts to dominate. On the 
 other hand, between the plateau and the nebular phase there is another transitional state, when an inflection 
 in the light curve can be seen (\S\ref{res:nick}). Hence the plateau duration is precisely 
 the time span between the secondary peak and the inflection. The inflection ($t_{i}$) in the $V$ band light 
 curve is observed to occur  nearly 103.2 days after the shock breakout. Hence the time interval between 
 the $t_{i}$ and $t_{p}$ is assumed as the duration of the plateau i.e. $(103.2-5.3) \sim 98$ days. 
 
 Once the hydrogen envelope is fully recombined and the ejecta becomes optically thin, the light curve  
 enters into the nebular phase and it is sustained mainly by the energy output from the 
 radioactive decays of the iron-group elements. During the plateau to nebular transition phase, the $V$-band magnitude 
 drops from $\sim$ 16.0 mag at around +90d (still in the plateau) to 18.1 mag 
 at +122d (in the exponential light curve tail), i.e. $\sim$ 2 mag in about one month. This drop is 
 remarkably smaller than that of SN 2005cs, but consistent with the 2-3 mag drop observed in normal 
 Type IIP SNe \citep{olivares10}. A linear fit to the tail from +120d to +400d gives the 
 following decline rates [in mag (100d)$^{-1}$]: $\gamma_B~\sim~0.44$, $\gamma_V~\sim~0.84$, 
 $\gamma_R~\sim~0.91$, $\gamma_I~\sim~1.03$ at $B, V, R,$ and  $I$ bands which are similar to the values found in 
 normal IIP SNe and comparable with the decay slope of $^{56}$Co to $^{56}$Fe, i.e. 0.98 mag (100d)$^{-1}$.


\section{Spectroscopic Evolution} \label{res:spec} 

\begin{figure*} 
\plotone{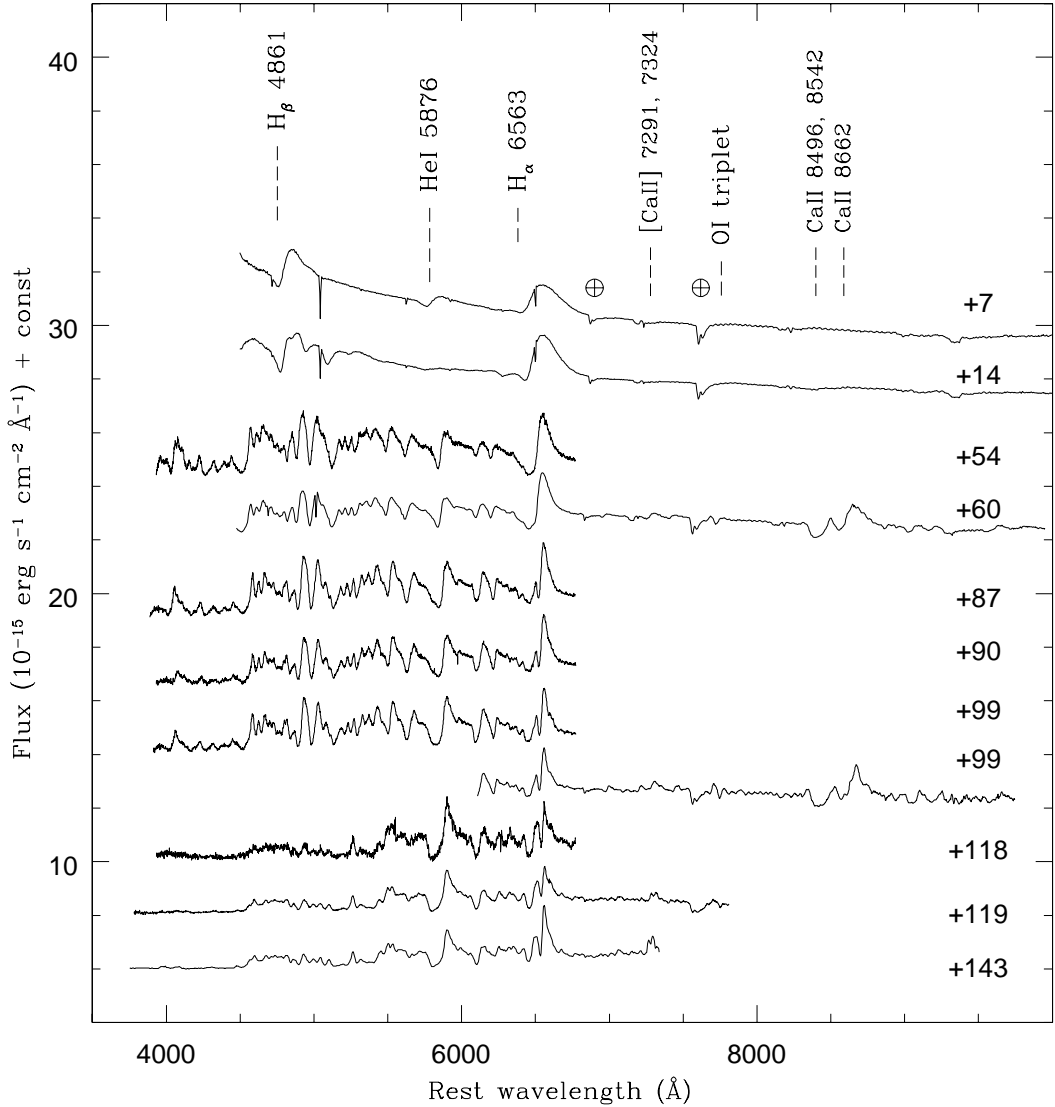} 
\caption{Doppler corrected flux spectra of \sn\  observed during +7 d to +143 d. 
         Prominent lines are marked. Telluric features are marked with $\earth$ symbol. 
         The sharp absorption dips seen near 5000 \AA\ and 6500 \AA\ in the +7d and +14d spectra are artifacts.
         The spectra signify the early photospheric phase (+7d,+14d), mid-plateau phase (+54d, +60d), 
         late-plateau phase (+87d, +90d, and +99d) and nebular phase (+118d, +119d, and +143d).} 
\label{fig:spec_all} 
\end{figure*}

 The spectra of \sn\ at 10 phases from +7d to +143d are shown in Figure \ref{fig:spec_all}. The spectra are 
 corrected for the recessional velocity of the host galaxy ($\sim$ $1567\pm3$ \kms\,)\footnote{http://leda.univ-lyon1.fr/}. 
 Spectral features are mainly identified as per previously published line identifications for IIP SNe 
 \citep{leonard02,pastorello04}. The two earliest spectra (+7d and +14d) show the blue continuum of blackbody emission and 
 have broad P-Cygni profiles of \ha\,, \hb\ and \Hei\ $\lambda$5876. The next two spectra (+54d and +60d) represent 
 the mid-plateau phase and are marked by a decrease in the continuum and the appearance of more number of P-Cygni profiles 
 for the permitted metallic (\Feii\,, \Scii,, \Tiii\,, \Baii\,, \Mgii\,), \Oi\ $\lambda$7773, \Nai\ D, and \Caii\ IR
 triplet lines similar to other normal Type IIP SNe (see \citealt{roy11} and references therein). The spectra at phases +87d,
 +90d and +99d represent the end stages of the plateau and these are marked by a redder continuum and decreasing line widths of
 hydrogen lines. The +118d, +119d and +143d spectra represent the nebular phase having negligible continuum and are marked by 
 pronounced emissions of forbidden lines \Caiia\, $\lambda\lambda$7291, 7323, and fading of the absorption features of 
 hydrogen and metallic lines.


\begin{figure*} 
\plotone{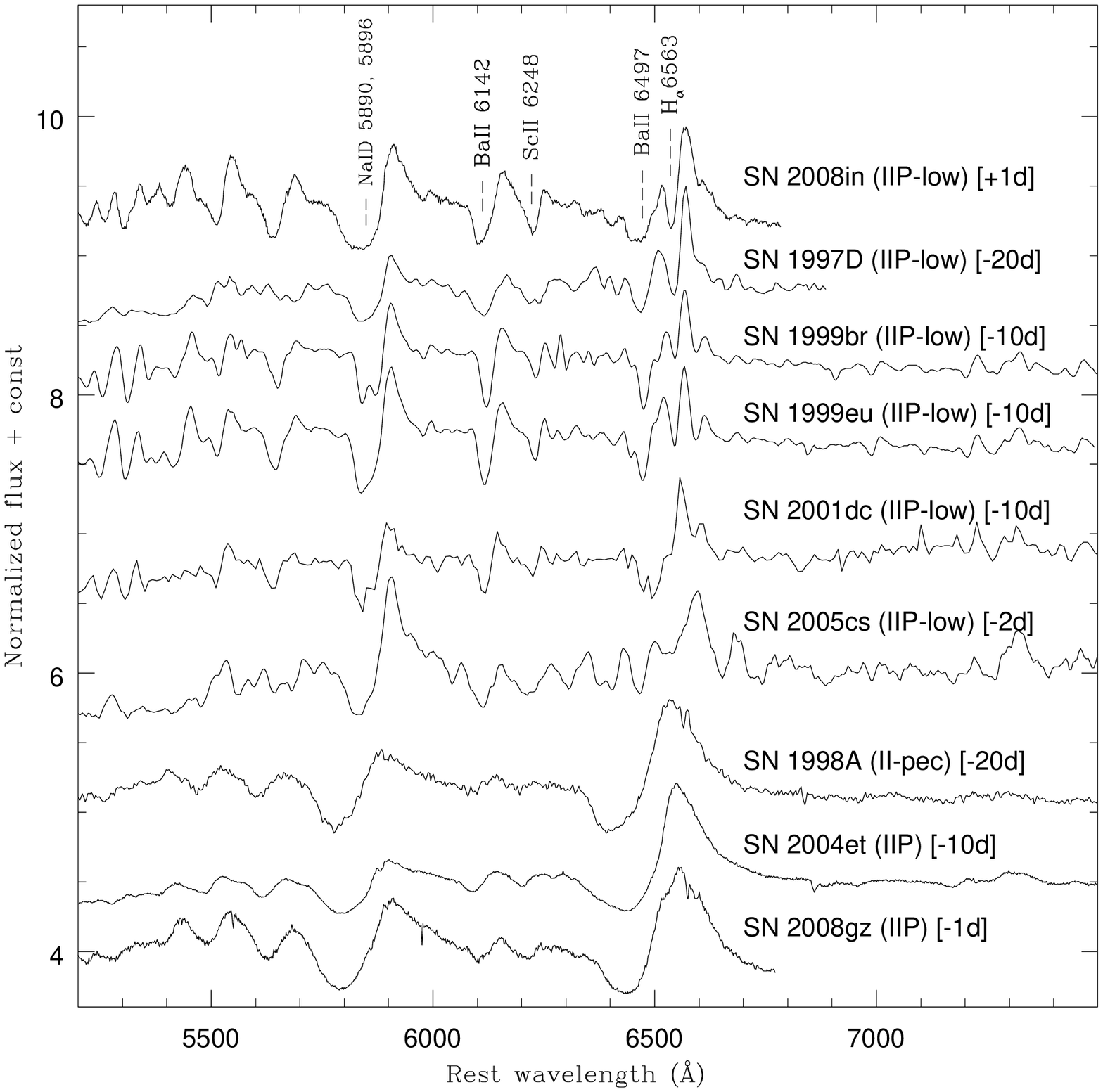} 
\caption{The end plateau spectrum (+99d) of \sn\,, is compared with five low-luminosity SNe 1997D, 1999br, 1999eu, 2001dc,
         2005cs (\citealt{pastorello09} and references therein); two normal SNe 2004et \citep{misra07}, 2008gz \citep{roy11}
         and a peculiar Type II SN 1998A \citep{pastorello05}, observed at comparable epochs. The phases quoted for each event
         are with reference to the  moment of inflection (t$_i$; see Figure \ref{fig:steep}), which marks the highest
         rate of decline in the $V$-band light curve during the end of the plateau and the beginning of nebular phase
         \citep{elmhamdi03b}.
         All the spectra have low spectral resolution ($\sim$ 10 \AA). The P-Cygni features of \Nai\ D, \Baii~$\lambda$ 6142, 
         \Scii~$\lambda$6248, \Baii~$\lambda$6497 and \ha\, are marked.} 
\label{fig:spec_ccsne} 
\end{figure*} 

\begin{figure*}
\includegraphics[width=15cm]{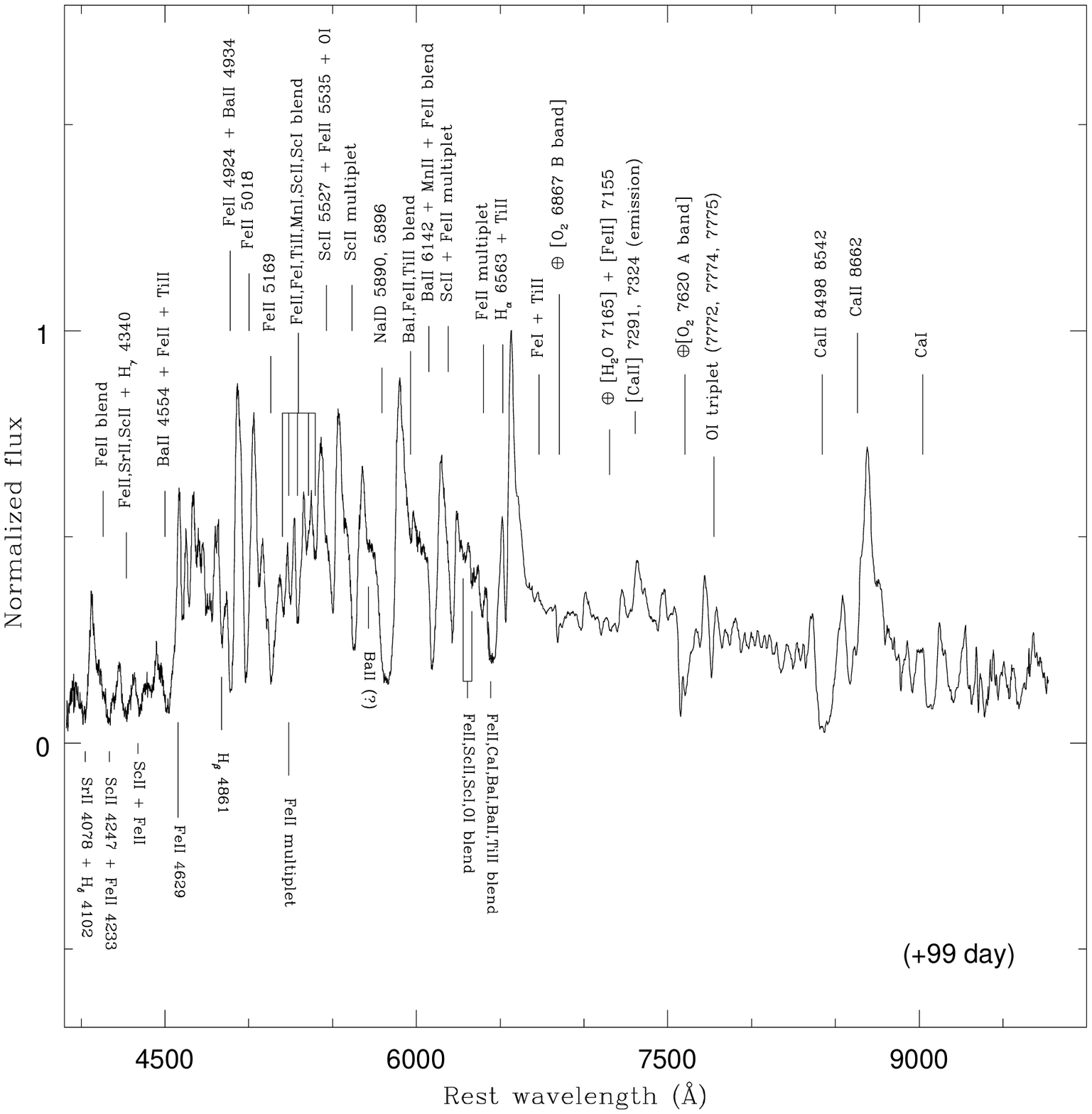}%
\caption{The spectral line identifications in a late-plateau phase (+99d) spectra are shown. Spectral features
          are mainly identified as per previously published line identifications for IIP SNe 
         \citep{leonard02,pastorello04}. Telluric features are marked with $\earth$ symbol.} 
\label{fig:line_id} 
\end{figure*}

 In Figure \ref{fig:spec_ccsne}, we compare the +99d spectrum of \sn\ with the spectra of other Type IIP SNe 
 observed at similar phases that is at roughly the time when the light curve changes from the photospheric plateau to the
 nebular phase. The phases quoted for each event are with reference to the moment of inflection ($ti$; see Figure
 \ref{fig:rotse}). The narrow P-Cygni \ha\ profile and the presence of strong lines of \Baii\ at $\lambda$6142 and
 $\lambda$6497 of \sn\ show striking resemblance to the low-luminosity SNe 1997D, 1999br, 1999eu, 2001dc and 2005cs. On
 the other hand, the normal Type IIP SNe 2004et, 2008gz show broader profiles of \ha\ and weaker lines of \Baii\,.
 It is noted that in the blue wing of \ha\ more metallic lines get resolved than 
 that in the normal IIP SNe and this arises due to smaller line widths of hydrogen lines seen in low-luminosity IIP SNe
 (\citealt{fraser10} and references therein). In addition, the \Baii\ line in the low-luminosity SNe show stronger
 absorption components due to lower ejecta temperature than that in the normal luminosity Type IIP SNe 2004et and 2008gz
 \citep{turatto98}. In Figure~\ref{fig:line_id}, we have identified the spectral features in a late-plateau 
 phase (+99d) spectrum which covers the full wavelength range from 0.4 to 0.95 \mum. For line identification we have 
 followed \citet{pastorello04}, where different spectral lines were identified for the +102d spectrum of a low-luminosity 
 SN 1999br. We are able to identify all the features and the spectral profiles of all the elements are similar to the 
 archetypal low-luminosity IIP SNe 1997D, 2005cs and 1999br (\citealt{pastorello04} and references therein). 


\begin{figure*} 
\includegraphics[width=15cm]{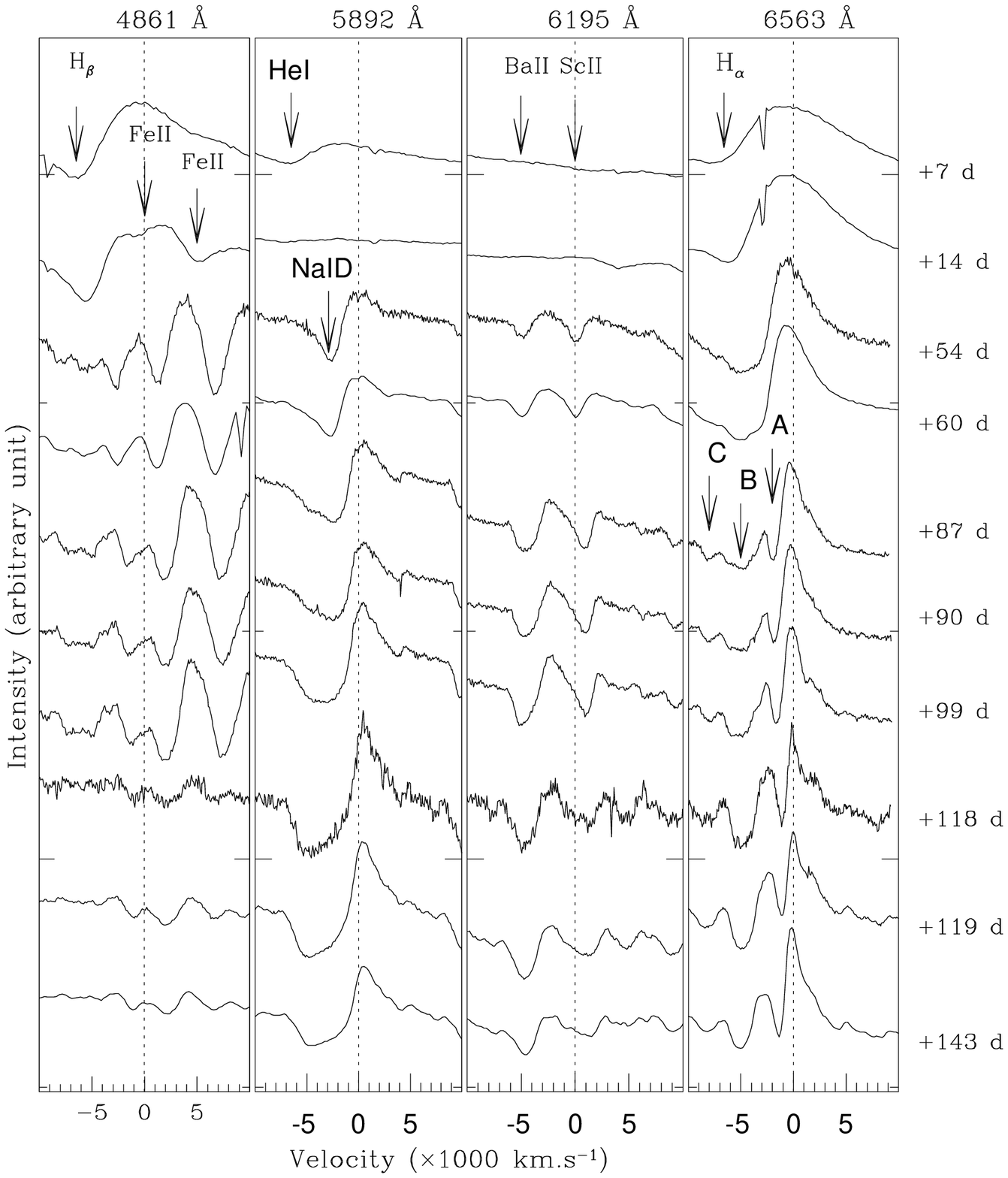}%
\caption{Spectral evolution of \hb\,, FeII $\lambda$4924, FeII $\lambda$5018, HeI $\lambda$5876, \Nai\ D $\lambda\lambda$5890, 5896,
         \Baii\ $\lambda$6148, \Scii\ $\lambda$6248, \Baii\ $\lambda$6497 and \ha\ lines of \sn\, during its transition from plateau 
          to nebular phase. The dotted line at zero velocity corresponds to the rest wavelength, marked at the top of each panel. 
         The sharp absorption dips seen in the \ha\ profile for +7d and +14d spectra are artefacts. } 
\label{fig:vel_line} 
\end{figure*} 
 
\begin{figure*} 
\includegraphics[width=9cm]{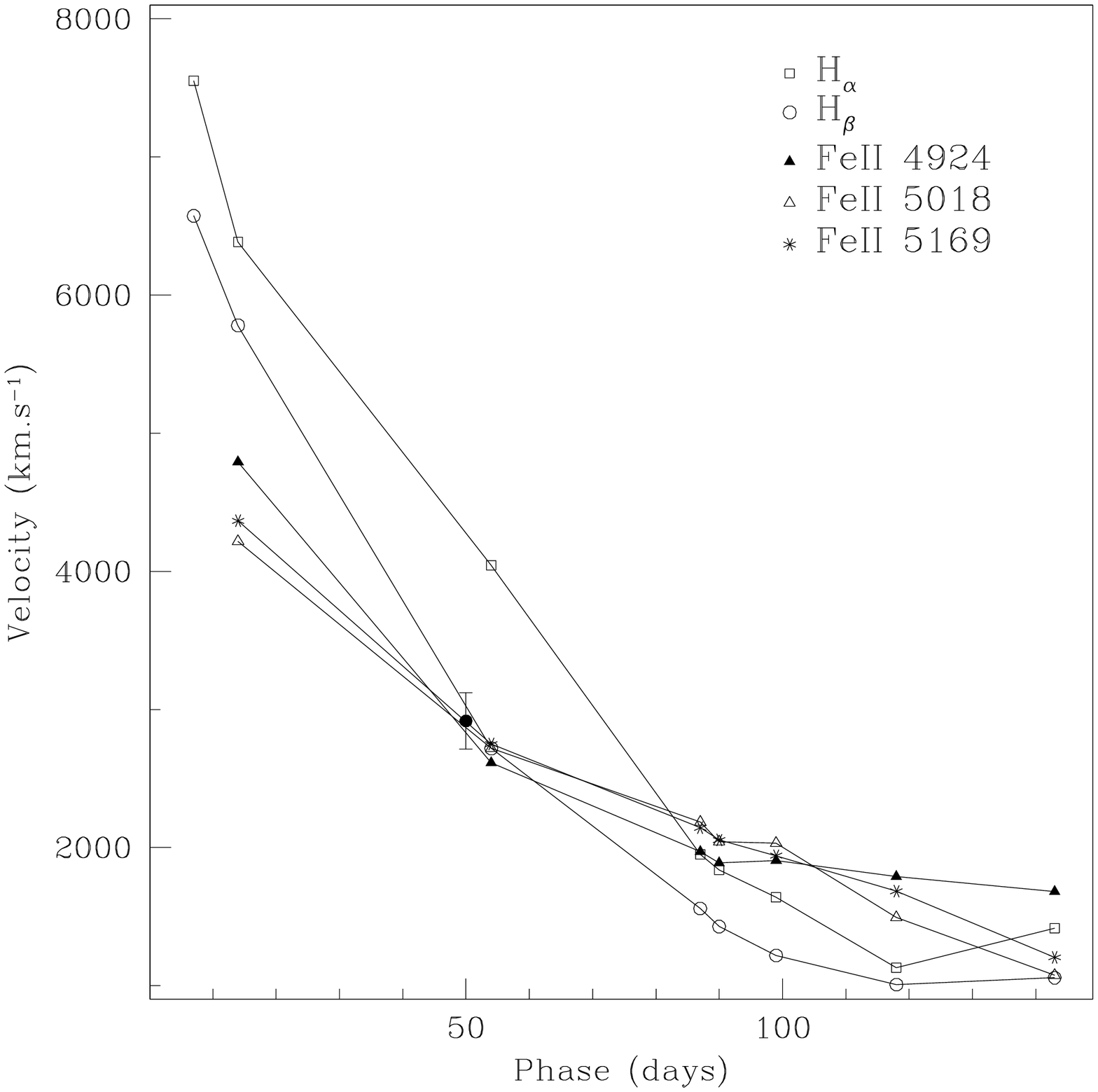}%
\includegraphics[width=9cm]{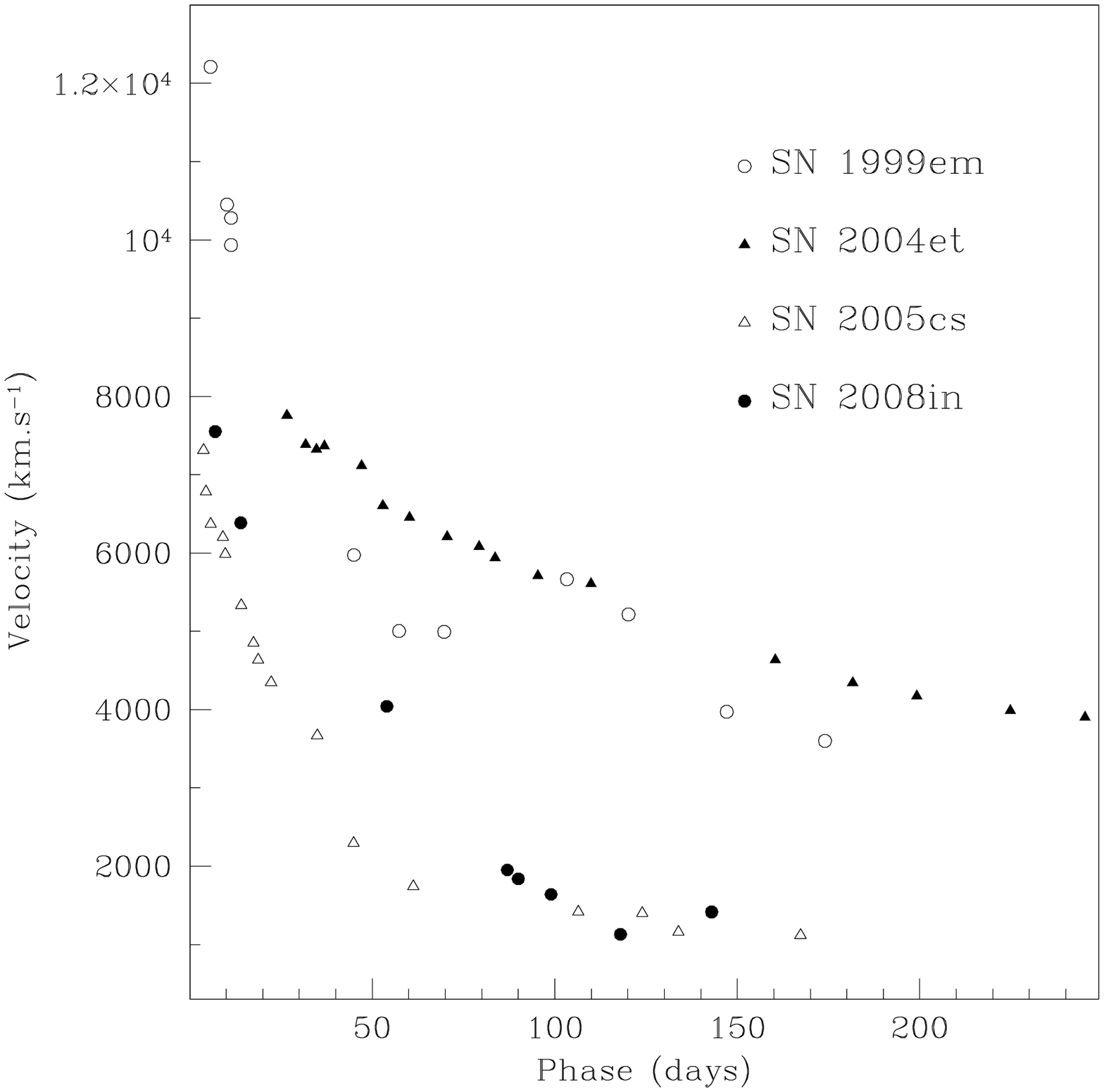}%
\caption{Evolution of expansion velocity of \sn\,. The left panel shows the velocity of different elements in the ejecta of \sn\,.
         The black filled circle shows the estimated value of mid-plateau photospheric velocity, calculated using the relation
         mentioned in \citet{roy11}. The right panel shows a comparison of velocity derived from \ha\ with that of low-luminosity
         event like SN 2005cs \citep{pastorello09} and normal Type IIP SNe 1999em \citep{elmhamdi03a} and 2004et \citep{sahu06}.}
\label{fig:vel_bulk} 
\end{figure*}

 Figure~\ref{fig:vel_line} shows the spectral evolution of \ha\,, \hb\,, \Nai\,D, \Feii\,, \Baii\,, and \Scii\ lines. In
 Figure~\ref{fig:vel_bulk} (left panel), we show the expansion velocities of the ejecta derived from Balmer and \Feii\
 lines. The later is a good indicator of photospheric velocity. The expansion velocities of the H-envelope are estimated using
 absorption minima of the P-Cygni profiles and at the two earliest phases (+7d and +14d) \ha\ and \hb\ show a broad 
 P-Cygni profile which, with time, becomes narrower at later phases keeping the position of the emission peak centered 
 near zero velocity. The \ha\ line velocity starts at about  7000 \kms\ at +10d, reaches 4000 \kms\ at +50d and flattens
 at a level of 1200 \kms\ in the nebular phases. It can be seen from the right panel of Figure~\ref{fig:vel_bulk} that
 in the comparable phases, the \ha\ line velocities of low-luminosity SNe are less than half those of normal IIP SNe 1999em and
 2004et, whereas the \sn\ velocities are more like low-luminosity SN 2005cs. To estimate the photospheric velocity of the transient
 we have computed the velocities of different \Feii\ lines $\lambda$4924, $\lambda$5018, and $\lambda$5169 at different phases.
 The first marginal detection of these lines is in the +14d spectrum and they became prominent in later stages of evolution.
 For \sn\,, the average velocity of these lines (and hence roughly the photospheric velocity) at +14d is about 4450 \kms\,,
 which is comparable with that of low luminosity SNe and much less than ordinary Type IIP events (see Figure 12 of
 \citealt{pastorello09}). 


\begin{figure*} 
\centering 
\includegraphics[width=15cm]{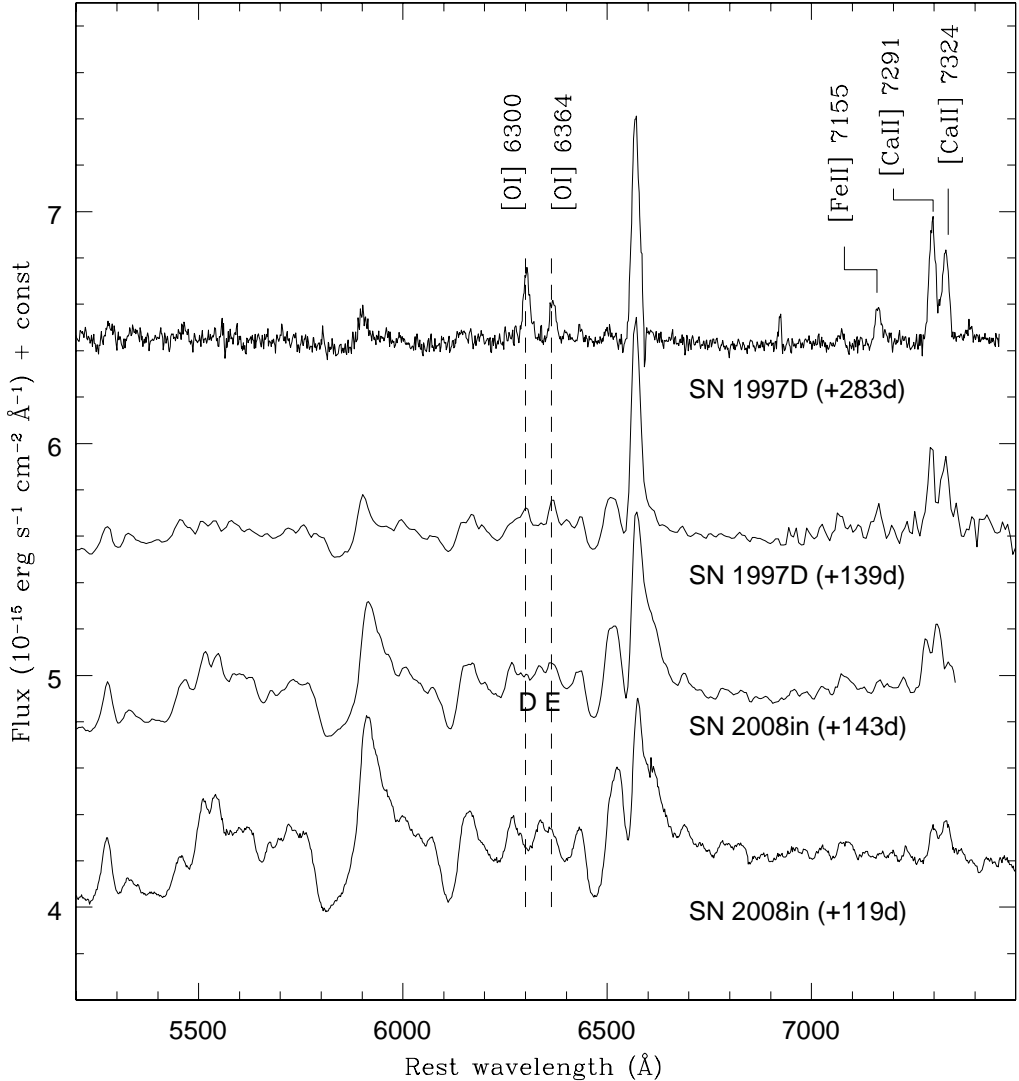}%
\caption{Forbidden lines during the nebular phases of \sn\, are compared with the low-luminosity SN 1997D. The lines [\Oi],
         [\Feii] and [\Caii] are marked.} 
\label{fig:spec_oi} 
\end{figure*} 

 The emergence of absorption dips in the blue wings of \ha\ is clearly seen in the post +60d spectra \footnote{This is the first
 time that the rapidly changing evolution of \ha\ profiles during the late plateau to nebular phase has been so densely covered in
 a low-luminosity SN} and the absorption dips have been marked with A, B and C in Figure 8. We distinguish `A' as an absorption dip
 due to \Tiii\ and \ha\ and its equivalent width (EW) increases from 12.85 \AA\ at +60d to 19.45 \AA\ at +143d. This progressively
 stronger absorption dip requires a steep deceleration of the \ha\ emitting zone along with its depletion by inner metal shells (like
 \Tiii\,). The broader absorption dip `B' is identified as a blend of \Baii\ $\lambda$6497, \Bai\,, \Cai\ and \Tiii. An exceptionally
 strong \Baii\ lines is quite typical for cooler ejecta of less energetic fain SNe. The segment `C' is speculated to be a footprint
 of \Feii\ multiplets as the evolution of its linewidths seems to be correlated with the other \Feii\ lines. It is noted, however,
 that in a few normal Type IIP SNe 1999em, 2004et and 2008gz, the spectral feature near 'C' at early plateau phases was also
 identified as a signature of high velocity \ha\ component, and in the present spectra its presence cannot be completely ruled out.

 The evolution of \hb\ is similar to the evolution of \ha. The P-Cygni profile of \hb\ is clearly visible in
 the +7d spectrum, and later on, the emission component is blanketed by various metal lines, mainly due 
 to \Feii\ $\lambda$4924, $\lambda$5018 and $\lambda$5169. Traces of \hb\ and \Feii\ 
 absorption dips indicate that from the beginning of the nebular phase H shells of ejecta and the regions 
 containing \Feii\ ions move with a comparable velocity, reaching an asymptotic value $\sim$ 1000 \kms\  nearly 
 140 days after the explosion. In the +14d spectrum, \Feii\ lines are marginally detected while in the spectra between  
 +60d and +99d they are prominent. We investigate the temporal change of EW of a relatively less blended \Feii\ $\lambda$4924
 and from the spectra normalized to the \ha\ peak value, we found that at an EW $\sim$ 0.67 \AA\, at +14d, increased 
 to 17.45 \AA\ at +54d and to 26.53 \AA\ in the +99d spectrum. Later on, the EW decreased to 19.07 \AA\ in the +118d
 spectrum and finally reaches to 18.27 \AA\ in the +143d spectrum. This rapid fall in the EW of inner metal-rich shells
 plausibly indicates a decrease in the opacity of the SN inner ejecta.

 Traces of the \Hei\ $\lambda$5876 line are also seen in the +7d spectrum (Figure \ref{fig:vel_line}). The ratio of EWs between
 \ha\ and \Hei\ for this spectrum is EW(\ha\,)/EW(\Hei\,) $\sim 1.99$. In the +14d spectrum, this ratio has increased 
 to 5.36. The steep decrement in EW of \Hei\ seems to be due to a quick recombination of \Hei\ ions as a result of the rapid 
 fall in temperature of the constantly rarefying ejecta. From +54d, the emerging P-Cygni feature of \Nai\ D becomes prominent. 
 This feature seems to be a perfect P-Cygni throughout our entire spectral sequence. It simply indicates an uniform spherical 
 distribution of \Nai\ in the SN ejecta. In the three high S/N spectra labeled with +54d, +60d and +87d, a tiny absorption dip
 overlaid on the emission component of the \Nai\ D P-Cygni is seen and since these spectra are redshift corrected, these dips
 are impressions of intervening interstellar matter present in the host galaxy. The absence of any similar absorption dip in
 the blue wing of the \Nai\,D profile, expected from the Galactic interstellar matter, confirms that there is little absorption
 from the Milky Way in the line of sight of \sn\,. 
  
 The spectral evolution of \Scii\ $\lambda$6248 and the s-process element \Baii\ $\lambda$6142 is clearly seen in \sn\,. 
 In the +7d and +14d spectra, there is no clear evidence for the presence of these two elements; however in the +54d
 spectrum they are prominent. The absorption components become stronger with time and persist until the +143d spectrum. In contrast,
 in the normal Type IIP SNe 2004et \citep{sahu06}, 1999em \citep{elmhamdi03a} and 2008gz \citep{roy11} these features disappeared
 at $\sim$ 170 day after the explosion. The ejecta of underluminous Type IIP SNe expand at a lower velocity than those of normal
 ones. So, the Ba lines in low luminosity events persist for a longer time just because the ejecta takes more time to
 cool-down. As a result it takes longer time to become optically thin. 
 
 In the nebular phase, the forbidden lines: [\Caii] $\lambda\lambda$7291, 7323, [\Oi] $\lambda\lambda$6300, 6364 and [\Feii]
 $\lambda$7155 are among the strongest features visible in the spectra. Also, the line ratios of emission lines of \Caii\ and the
 \Oi\ doublet in late nebular phases is a good indicator of the progenitor mass \citep{fransson87}. In Figure~\ref{fig:spec_oi}, we
 compare the +139d and +283d spectra of SN 19997D (Benetti et al. 2001) with the nebular phase spectra of \sn. The [\Caii] doublet
 is seen in the +99d, +119d and +143d spectra of \sn, whereas a small footprints of the \Oi\ doublet can be seen in the +143d spectrum.
 Presence of [\Feii] line is not seen in our spectra. Considering that the explosion epoch of SN 1997D was quite uncertain, it is
 likely that the spectral evolution of \ha\ and forbidden lines of \sn\ is quite similar to that of SN 1997D.


\section{extinction and distance toward \sn\,} \label{res:DistExt} 

 In order to determine the bolometric light curve and other physical parameters, a correct estimate of the extinction
 and distance toward the SN is essential. We adopt the Galactic reddening along the line-of-sight of \sn\, as derived 
 from the 100 \mum\, all sky dust extinction map \citep{schlegel98}, i.e.  \ebv\, = $0.0224\pm0.0003$ mag. In order
 to estimate reddening due to the host galaxy M61, we used the spectrum of \sn\ taken on 2009 February 17 from the 2-m IGO 
 telescope having 
 good S/N ($\sim$ 40) and corrected for mean heliocentric radial velocity of the host ($cz \approx 1567\pm3$ \kms\,).
 Near the zero velocity, the spectrum showed a tiny absorption feature overlaid on the emission component of the P-Cygni profile 
 of \Nai\,D (\S~\ref{res:spec}). The EW of this absorption feature was computed as 0.535$\pm$0.713 \AA. The error in EW is 
 estimated using Equation 6 of \citet{vollmann06}. It is known that the EW of the interstellar absorption bands is well 
 correlated with the reddening \ebv\ estimated from the tail of SN Ia color curve \citep{barbon90, richmond94, turatto03} 
 and hence using the empirical relation \ebv\, $=$ $-0.01 + 0.16$EW (EW in \AA), given by \citet{turatto03}, we obtain the
 host contribution as \ebv\, $\approx 0.076\pm0.104$, which is considerably higher than the Galactic contribution. This is
 consistent with the absence of any absorption feature due to Galactic interstellar matter in the blue wing of the emission
 component of \Nai\,D profile. Consequently, we adopt \ebv\, (estimated as a sum of Galactic and host galaxy reddening) of
 0.0984$\pm$0.104 mag for \sn. This corresponds to a visual extinction ($A_V$) of $0.305\pm0.322$. Considering the uncertainty in
 the \ebv\ estimated using \Nai D lines, a lower value of \ebv\ of 0.0448$\pm$0.0006 mag (twice the Galactic reddening) and the
 corresponding $A_V$ of $0.139\pm0.002$ cannot be ruled out. We will discuss its implication for the derived properties of the SN.

 The Hubble flow distance of the host galaxy M 61, after correction for the Virgo infall, is estimated as 13.7$\pm$1.1 Mpc
\footnote{The cosmological model with $H_0$ = 70 \kms\, Mpc$^{-1}$,$\Omega_{m}$ = 0.3 and $\Omega_{\Lambda}$ = 0.7 is assumed
 throughout the paper and the uncertainty corresponds to a local cosmic thermal velocity of 208 \kms \citep{terry02}.}. The
 distance estimated through the Tully-Fisher method is $12.1\pm2.7$ Mpc\footnote{http://nedwww.ipac.caltech.edu/}. 
 Additionally, we also calculated the distance following the Standard Candle Method for Type II SNe \citep{hamuy02, hamuy05, hendry05}.
 It is found that there is a strong correlation between the distance of Type IIP SNe along with their mid-plateau apparent
 $V$ and $I$ band magnitudes and the mid-plateau photospheric velocity, for a given cosmological model. For \sn\, we estimate the
 mid-plateau ($\sim +50$d) apparent magnitudes $\sim 15.56\pm0.05$ mag for $V$ and $\sim 14.77\pm0.06$ mag for the $I$ band.
 Considering the $V$ and $I$ band extinctions toward the SN, $0.305\pm0.322$ mag and $0.183\pm0.193$ mag respectively,
 we derive a distance of $\sim 12.23\pm1.87$ Mpc. The adopted value of mid-plateau photospheric velocity is $2694.67\pm70$ \kms
 (see \S\ref{res:proge}). Combining the above three measurements, we adopt the weighted
 mean distance of $13.19\pm1.09$ Mpc, which corresponds to a distance modulus of $30.6\pm0.2$. 
 

\begin{figure*} 
\centering 
\includegraphics[width=15cm]{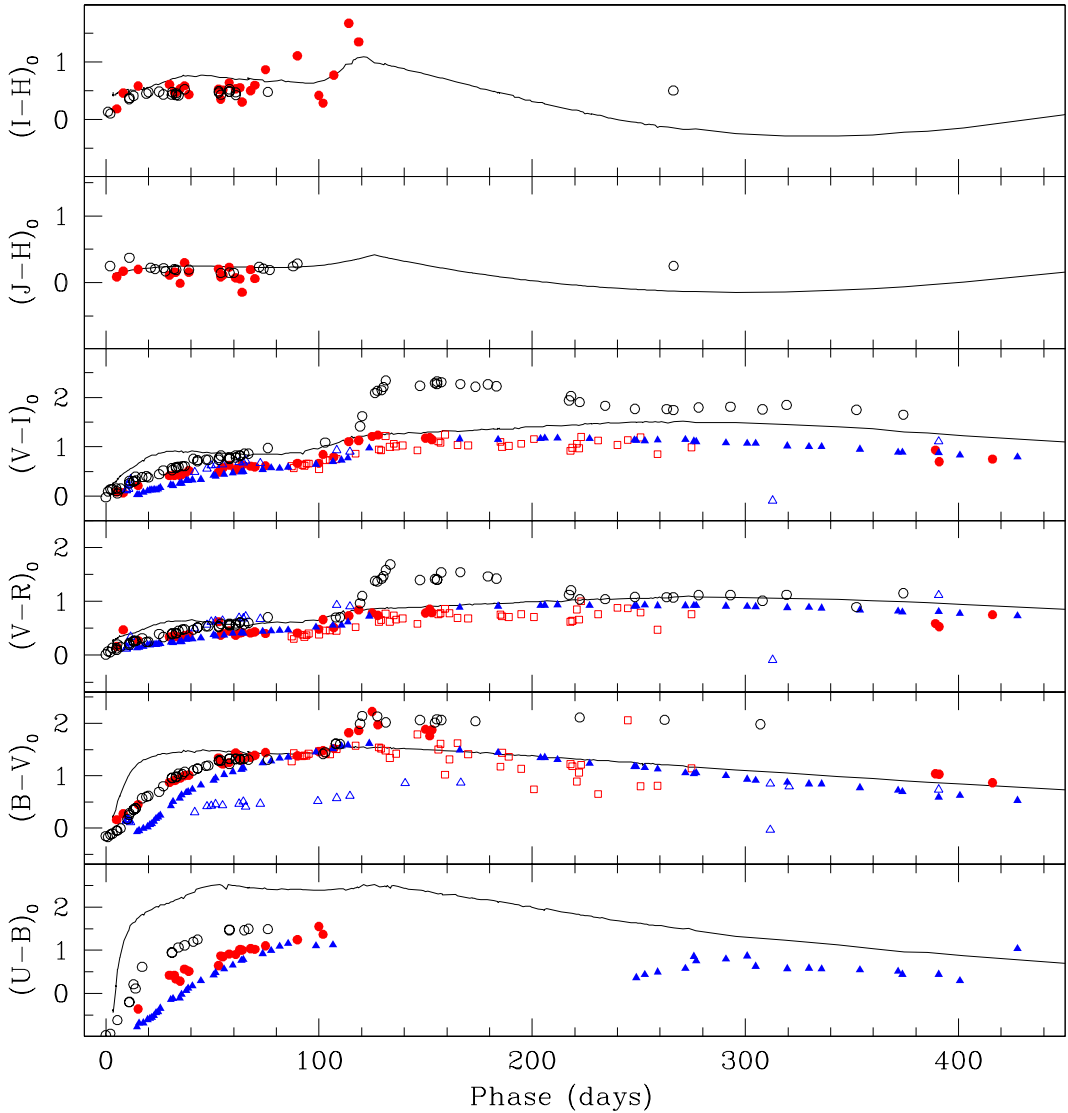}%
\caption{Colour curve of \sn\ is shown with filled circles. Also shown are the SNe 1987A (solid line), 
 1999em (open triangle), 2004et (filled triangle), 2005cs (open circle) and 2008gz (open square).} 
\label{fig:colorcur} 
\end{figure*} 


\section{Color Evolution and Bolometric Flux} \label{res:ColBol} 

 Figure~\ref{fig:colorcur} shows the temporal evolution of the reddening-corrected broadband colors of \sn. For
 comparison, we overplot the color curves for well studied SNe 1987A \citep{suntzeff90}, 1999em \citep{elmhamdi03a},
 2004et \citep{sahu06, maguire10}, 2005cs \citep{pastorello09} and SN 2008gz \citep{roy11}. After small differences
 during the initial phases, the plateau phase color evolution of 
 all Type II SNe is more or less similar. The $(U-B)_{0}$ and $(B-V)_{0}$ colors are blue during early photospheric phases and they 
 become redder by about 1-2 mag in the plateau phase while the $(V-R)_{0}$ and $(V-I)_{0}$ colors evolves slowly and become red only 
 by about 0.5 mag. The $(J-H)_{0}$ color remains constant at $\sim$ 0.25 mag. During the transition phase from plateau to nebular, the 
 low luminosity SN 2005cs showed a striking red peak in the $(B-V)_0$, $(V-R)_0$ and $(V-I)_0$ colors. For \sn\,, this red peak 
 is not present and its color evolution is found to be consistent with the normal Type IIP SNe. 

\begin{figure*} 
\centering 
\includegraphics[width=15cm]{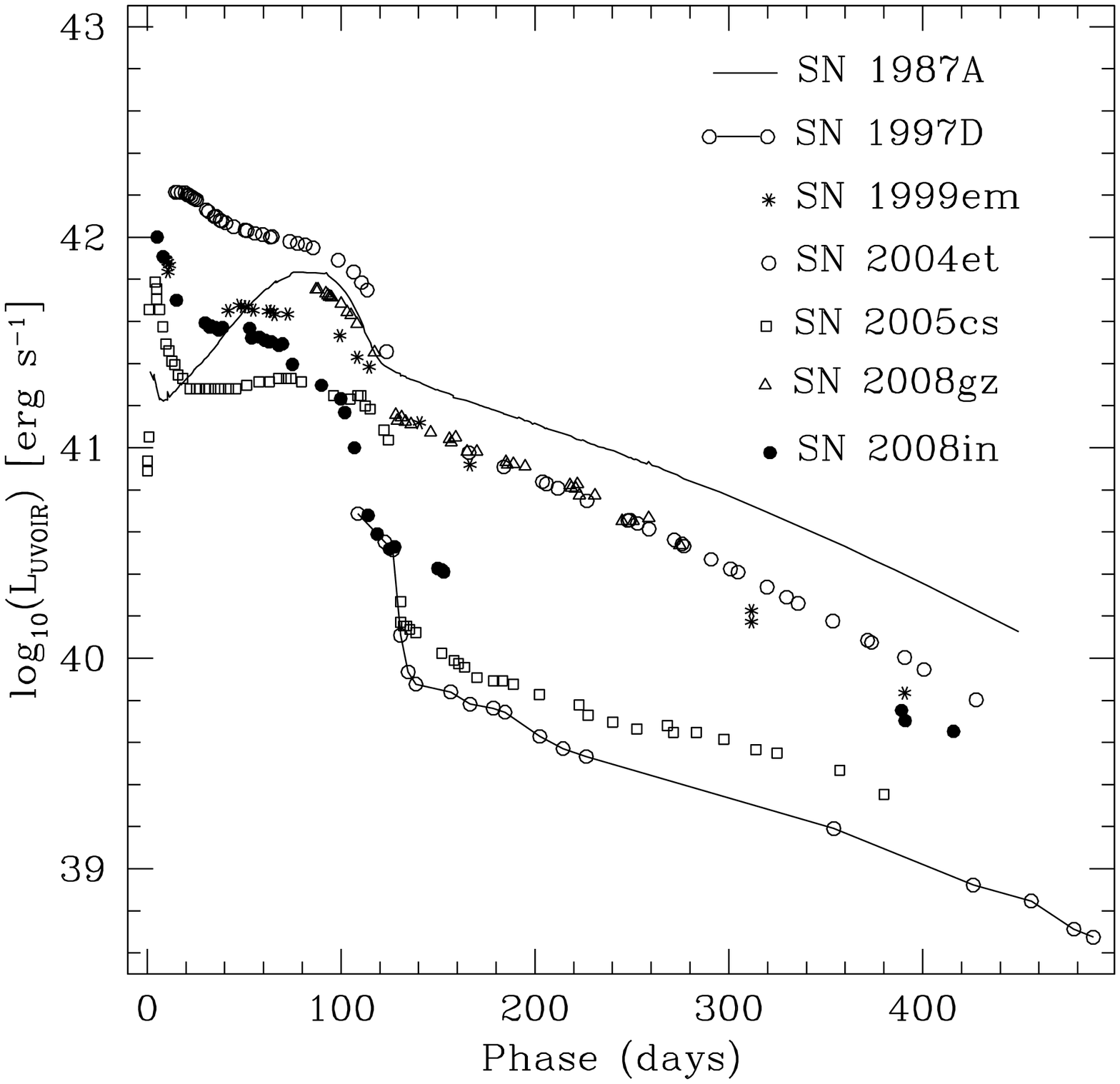}%
\caption{Comparison of the quasi-bolometric light curve of \sn\, with the low-luminosity SNe 1997D, 2005cs;
         the normal SNe 1999em, 2004et, 2008gz; and the peculiar Type II SN 1987A. } 
\label{fig:bolcur} 
\end{figure*}

 The quasi-bolometric light curve estimated from the UV, optical and IR broadband ($UVOIR$) magnitudes of \sn\ is shown in 
 Figure~\ref{fig:bolcur} along with those of other Type II SNe. The extinction-corrected magnitudes are first converted into
 fluxes using zero-points given by \citet{bessell98} and then the total flux in $UVOIR$ bands is obtained after a linear
 interpolation and integration between 0.203 and 1.67 $\mum$. The fluxes were calculated on those nights for which we had complete  
 observations in $UBVRI$ bands. For the initial two weeks the SN was detected by {\it Swift}/UVOT in near ultraviolet bands
 with a significant flux density. This contribution has been accounted for while measuring the net $UVOIR$ flux.
 It has been assumed that contribution in the $U$ band is mainly important during the plateau and 
 decreases rapidly to about 5\% in the nebular phase (see \citealt{misra07, roy11} and references therein). From  the
 light curve it is noticeable that the near ultraviolet flux is almost negligible beyond 20 days after explosion.
 The $JH$ contribution during the plateau is calculated from our data, whereas for the nebular phase, due to lack of data,
 we are not able to make any direct measurement. For most low luminosity SNe IIP, the NIR flux contribution in the nebular phase 
 is about 50\% of the total flux \citep{pastorello09}. We have therefore increased the net flux by 55\% to account for the
 maximal contributions from the $U$ and NIR bands at phases later than +140d. Figure \ref{fig:spec_ccsne} and \ref{fig:bolcur}
 clearly demonstrate that spectroscopically \sn\ appears to be like low-luminosity SNe IIP, but photometrically it appears quite
 normal. In the following section, the $UVOIR$ curve is used to estimate the amount of radioactive $^{56}$Ni and other physical
 parameters that characterize the explosion and the progenitor star.  


\section{Physical Parameters} \label{res:parameter} 

 The physical entities that seem to govern the entire scenario are mainly associated with the nature of the progenitor 
 and the radioactive elements (mainly $^{56}$Ni), generated inside the inner portion of the ejecta during the explosion. 
 $^{56}$Ni is synthesized by the explosive burning of Si and O during the shock breakout \citep{arnett80, arnett96}. Over time
 this material is eventually converted to $^{56}$Co and then to $^{56}$Fe by means of radioactive transitions having decay times
 of 8.77 and 111.3 days respectively. The $\gamma$-rays and neutrinos emitted during this process sustain the nebular phase
 light curve and consequently, the observed tail luminosity becomes a good tracer for the ejected synthesized $^{56}$Ni.


\subsection{Produced Radioactive Nickel} \label{res:nick} 

 We use different methods to estimate mass of $^{56}$Ni. \citet{hamuy03} proposed a relation between bolometric tail luminosity and 
 the synthesized $^{56}$Ni during core collapse SNe, by considering the underlying assumption that all $\gamma$-rays emitted during the 
 radioactive decay make the ejecta thermalised. For \sn\,, the time interval spanned by the observations is about 416 days,
 where the first $\sim$ 100 days are reserved for photospheric evolutionary processes. The average $V$ band magnitude during 
 the nebular phase calculated using the data between +114d and +416d is $\sim$ 18.91, which corresponds to the $V$ magnitude at $\sim$    
 +222d. Taking the extinction correction as ($A_V = 0.305\pm0.322$ mag; \S\ref{res:DistExt}), a bolometric 
 correction of $0.26 \pm 0.06$ mag \citep{hamuy01} and a distance modulus $30.6\pm0.2$, the derived tail luminosity at this fiducial 
 time is $(3.02\pm1.95)\times10^{40}\,{\rm erg\,s^{-1}}$. Within the errors, this value is consistent with the bolometric flux at  
 a comparable epoch, determined in \S\ref{res:ColBol}. This implies that the amount of $^{56}$Ni produced is this process 
 is $M_{\rm Ni} =0.0157\pm0.0102$\msun. 
 
\begin{figure}
\includegraphics[height=13cm]{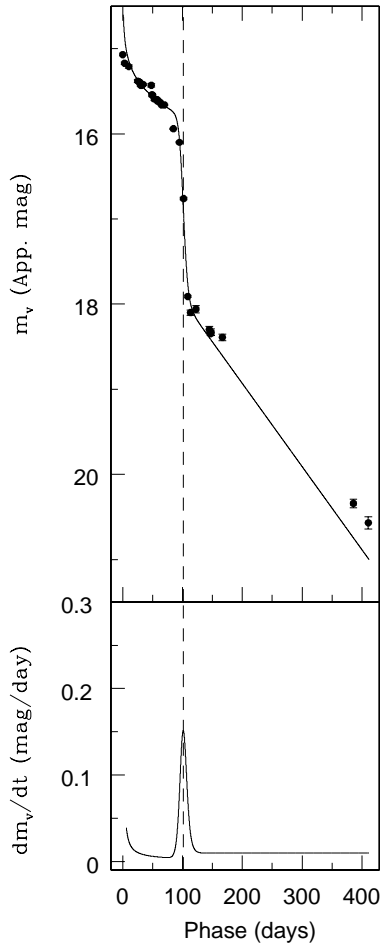}%
\caption{Determination of steepness parameter from the apparent $V$-band light curve of \sn. See text for details.} 
\label{fig:steep} 
\end{figure}

 All Type IIP SNe show an inflection in the light curve during the transition from the plateau to the nebular phase. 
 Statistically it has been shown that the steepness of the $V$-band light curve slope (defined as S=d$m_{\rm V}$/dt) at 
 the inflection time ($t_{i}$) is anti-correlated with \nickel\, mass \citep{elmhamdi03b}. For \sn\,, the $V$ band 
 light curve with its well-sampled transition phase, shows a value of steepness ${\rm S}=0.151\pm0.044$ mag d$^{-1}$ 
 (Figure~\ref{fig:steep}) and the epoch of inflection is $t_{i} \approx$ +101.5d with respect to date of discovery. This
 corresponds to $M_{\rm Ni}=0.0175\pm0.002$\msun. This result is consistent with the value measured using the \citet{hamuy03}
 scheme. According to \citet{elmhamdi03b}, the amount of $^{56}$Ni can also be derived from the plateau absolute $V$-band magnitude
 using the relation ${\rm log} M_{\rm Ni} = -0.438 M_{V}(t_{i}-35)-8.46$. Here $M_{V}(t_{i}-35)$ is the absolute $V$ magnitude 
 35 days prior to the day of inflection. For \sn\,, $M_{V}(t_{i}-35) \approx -15.23$, which again corresponds to a 
 $^{56}$Ni mass around 0.016\msun.  
  
 Comparison of the tail luminosity with that of SN 1987A $-$ a well studied proximate event, is also used for estimation of the 
 $^{56}$Ni mass. A linear least square fit on the nebular light curve tail shows that at +222d the luminosity 
 of \sn\ is about 1.54$\times10^{40}\,{\rm erg\,s^{-1}}$, while SN 1987A had a luminosity 
 $\sim 1.10\times10^{41}\,{\rm erg\,s^{-1}}$ (Figure \ref{fig:bolcur}). Since the $^{56}$Ni mass produced by SN 1987A 
 is about 0.075\msun, the amount of $^{56}$Ni in the case of \sn\ is $[(1.54/1.10)\times10^{-1}]\times0.075 \approx 0.0105$\msun. 
 
 The above estimates are consistent with each other and hence we adopt a mean value for $^{56}$Ni mass as $0.015\pm0.003$\msun\,


\subsection{Explosion Energy and Mass of Progenitor Star} \label{res:proge} 

 We use the radiation-hydrodynamical simulations of core-collapse IIP SNe by \citet{dessart10} to infer the explosion 
 energy ($E_{0}$ - kinetic plus thermal, expressed hereafter in units of $10^{51}$ erg or foe). These simulations suggest 
 that in a given progenitor larger explosion energies yield larger ejecta velocities and this implies that the ejecta kinematics
 can be used to put constraint on $E_{0}$. For \sn, the expansion rate of the H-rich progenitor envelope as derived from absorption
 minima in \ha\ at 15d after shock breakout is about 6300 \kms\ (see Figure~\ref{fig:vel_line}), which is higher than 4700 \kms\
 (for SN 2005cs; \citealt{pastorello09}) and lower than 8800 \kms (for SN 1999em; \citealt{elmhamdi03a}). 
 For the SNe 2005cs and 1999em, the simulation results (for explosion of a non-rotating solar metallicity pre-SN stars)
 predicted $E_{0} \sim$ 0.3 foe and $\ga$ 1 foe respectively and these values are found to be consistent with that determined from 
 the actual hydrodynamical modeling of the SN light curves, e.g. 0.4 foe \citep{utrobin08} and 0.3 foe \citep{pastorello09} 
 for SN 2005cs; 1.3 foe \citep{utrobin07a} and 1.25 foe \citep{bersten11} for SN 1999em. For \sn\,, the simulations suggest $E_{0}$
 $\sim$ 0.5 foe. 

 Accurate determinations of explosion parameters such as $E_{0}$, the ejected mass ($M_{\rm ej}$) and the pre-SN radius ($R_{0}$) of
 the progenitor require detailed hydrodynamical modeling of the light curves and spectra and this being a non-trivial task (beyond the
 scope of this paper) has only been attempted for a few SNe. In order to have an estimate of explosion parameters of \sn\, here, we
 employ the analytical relations derived by \citet{litvinova85} correlating the observed parameters ($M_{Vmp}$ - the mid-plateau 
 absolute magnitude at $V$, $v_{mp}$ - the mid-plateau photospheric velocity and $\Delta t_{p}$ -the plateau duration) with the physical 
 parameters ($E_{0}$, $M_{\rm ej}$ and $R_{0}$) based on a grid of hydrodynamical models for different values of physical parameter
 for Type IIP SNe. We note, however, that these approximate formulae have limitations owing to the poorly-measured observables
 and the simplified physical conditions such as non-inclusion of the effect of nickel heating, use of old opacity tables, neglecting 
 the effect of line opacity and using outdated pre-SN models (\citealt{smartt09, bersten11} and references therein). As a result, 
 only approximate values of the physical parameters can be derived using these relations. Fortunately, the observed parameters
 are derived very accurately for \sn. The $\Delta t_{p}$ is $\sim 98$ days (\S\ref{res:PlatNeb}), the $v_{\rm mp}$ is
 $2694\pm70$\footnote{This is the mean value of the velocities computed from the lines of \Feii\, $\lambda$4924, $\lambda$5018
 and $\lambda$5169 in the +54d spectrum.} \kms\, and the $M_{Vmp}$ is estimated as $-15.32\pm0.38$ mag\footnote{These observed
 values of $M_{Vmp}$ and $v_{mp}$ can be compared with the estimates derived empirically between the mass of $^{56}$Ni, and the
 $v_{mp}$ and $M_{Vmp}$ for a larger sample of IIP SNe (see Section 8.3 of \citealt{roy11}). Using the $^{56}$Ni mass of
 $0.015\pm0.003$\msun\, we find $v_{mp} =2916\pm 220\kms\,$ and $M_{Vmp} = -15.37\pm0.23$ mag which are consistent with those
 measured observationally.} Now, employing analytical relations, we estimate $E_{0}$ $\sim$ 0.54 foe, $M_{\rm ej} \sim 16.7$
 \msun\ and $R_{0}$ $\sim 127 R_{\odot}$. The explosion energy derived in this way is consistent with that predicted from the
 hydrodynamical simulations of the ejecta kinematics. 

 The explosion energy of \sn\ indicates that the event was less energetic than the standard IIP SNe 1999em, 2004et and more
 energetic than SN 2005cs. Assuming a net mass loss $\sim 0.5$M$_\odot$ due to stellar wind \footnote{\sn\ was not detected
 in X-ray. The lack of X-ray emission can be used to constrain the mass-loss rate of the progenitor system that could be
 heated by the outgoing shock to X-ray emitting temperatures. Following the discussion in \citet{immler07} and references
 therein, an upper limit to the pre-SN mass-loss rate of
 $5 \times 10^{-6}~M_{\odot}~{\rm yr}^{-1}~(v_{\rm w}/10~{\rm km~s}^{-1})$ with an uncertainty of a factor of two to three
 is obtained.} and accounting for a compact remnant with mass $\sim 1.5-2$M$_\odot$ \citep{sahu06}, we find that the 
 initial mass of the progenitor to be $\leq 20$M$_\odot$.  
 

\begin{figure*}
\includegraphics[width=15cm]{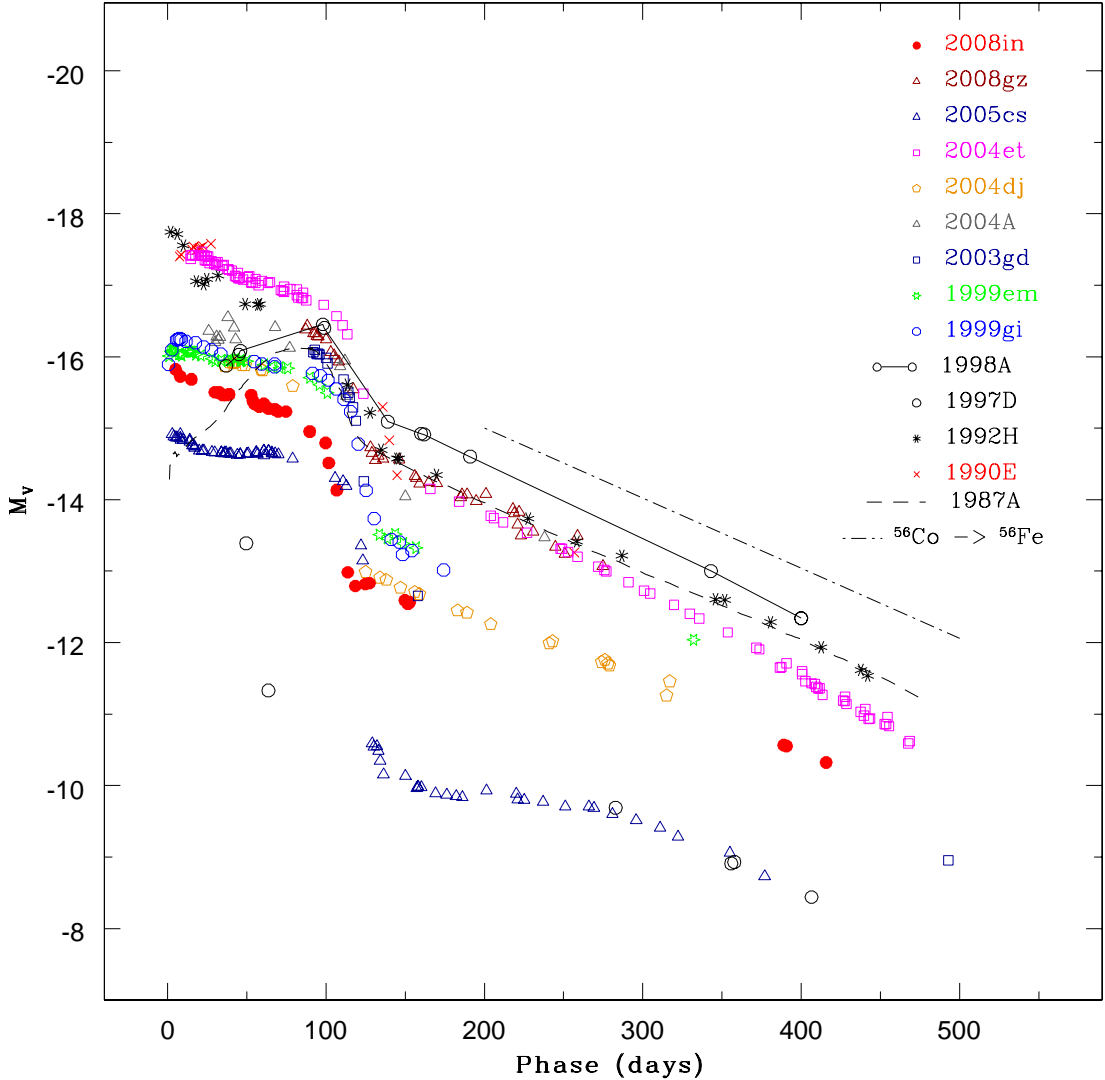}%
\caption{A comparison of the absolute $V$-band light curve of \sn\ with the low-luminosity SNe 1997D, 2005cs;
         the normal SNe 1999em, 2004et, 2008gz, 2004dj, 2004A, 2003gd, 19992H, 1999E, 1999gi; and the 
         peculiar Type II SNe 1987A, 1998A. The decline rate of emission expteced from radioactive decay of 
         $^{56}$Ni to $^{56}$Co to $^{56}$Fe is shown with dot-dashed line.} 
\label{fig:abs_lc} 
\end{figure*} 
 

\section{Discussion in context of other Type IIP SNe} \label{res:comp}

 In Figure \ref{fig:abs_lc}, we compare the light curve of \sn\ in absolute $V$-band magnitude with a sample of 13 other well
 studied Type II SNe taken from the literature (see \citealt{misra07} and \citealt{roy11} for the references).
 The sample includes two low-luminosity Type IIP SNe 1997D, 2005cs; nine normal Type IIP SNe 1990E, 1992H, 1999gi, 1999em, 2003gd, 2004A, 
 2004dj, 2004et, 2008gz and two peculiar Type II SNe 1987A, 1998A. It is seen that the Type II SNe show a wide range of mid-plateau 
 luminosity, i.e. from $-14$ to $-17$ mag. The data on low luminosity SNe (see also Table 4 in \citealt{pastorello06})
 indicate that their $M_{Vmp}$ lie between $-14$ and $-15$ mag, with the only exception of SN 2002gd, whereas the other SNe are
 seen to lie between $-16$ and $-17$ mag. With the $M_{Vmp}$ of $-15.32$ mag\footnote{Considering a lower value of
 $A_V = 0.139\pm0.002$, we get $M_{Vmp} = -15.1\pm0.2$}, the \sn\ present another case which occupies the gap between a low and 
 normal luminosity Type IIP events. 

 The tail luminosity is an important indication of ejected nickel mass. The average tail luminosity of \sn\ (Mv $\sim -12.16$) 
 is nearly 2.3 mag brighter than those of the low luminosity SNe 1997D and 2005cs ($\sim -9.6$ mag), and about 1.3 mag fainter
 than the normal Type IIP SNe 1992H, 2004et and 2008gz ($\sim -13.5$ mag). A close inspection shows that the average tail luminosity
 of \sn\ is roughly similar to Type IIP SNe 2004dj in NGC 2403 and 2003gd in M74, whereas their plateau is brighter ($\sim 0.5-0.8$
 mag at mid-plateau) than \sn\,. These comparison of nebular phase luminosities with other SNe are consistent with a quite modest
 radioactive $^{56}$Ni production for \sn\ (\S\ref{res:nick}).

 The luminosity and shape of the tail depend on the $^{56}$Ni mass and the radiant energy 
 per unit ejected mass ($E/M_{\rm ej}$). The first parameter determines the absolute magnitude, while the second determines its 
 decay rate (see \citealt{turatto98} and references therein). The measured value of ejected $^{56}$Ni mass for 
 SNe 2003gd, 2004dj and 2008in are respectively $\sim 0.016$ \citep{hendry05}, 0.017 \citep{venko06} 
 and 0.015\msun\,. These are more than twice the $^{56}$Ni amount synthesized by the low-luminosity 
 SNe 1997D ($\sim 0.002$\msun\,, \citealt{turatto98}) and 2005cs ($0.003-0.004$\msun\,, \citealt{pastorello09}), 
 although at least three times less than the $^{56}$Ni produced by normal Type IIP SNe such as 
 1992H ($\sim 0.075$\msun\,, \citealt{clocchiatti96}), 2004et ($\sim 0.06$\msun\,, 
 \citealt{sahu06, maguire10}) and 2008gz ($\sim 0.05$\msun\,, \citealt{roy11}). Similarly, the measured 
 value of $E/M_{\rm ej}$ ratio for \sn\ is  larger than that for the low-luminosity SN 1997D and lower than other 
 Type IIP events
 \footnote{ The measured value of the $E/M$ ratio for \sn\ is  
 $3.5\times\,10^{49}$erg~\msun\,$^{-1}$, which is twice the E/M ratio measured for SNe 1997D 
 ($1.7\times\,10^{49}$erg~\msun\,$^{-1}$, \citealt{turatto98}), but it is half  those of SN 2008gz 
 ($7.3\times\,10^{49}$erg~\msun\,$^{-1}$, \citealt{roy11}), SN 2004dj 
 ($7.3\times\,10^{49}$erg~\msun\,$^{-1}$, \citealt{venko06}) and SN 2004et 
 ($7.5\times\,10^{49}$erg~\msun\,$^{-1}$, \citealt{sahu06}). Finally, it is barely a third the E/M 
 ratios of SNe 1987A ($11.3\times\,10^{49}$erg~\msun\,$^{-1}$, \citealt{hamuy03}) and 2003gd  
 ($13.0\times\,10^{49}$erg~\msun\,$^{-1}$, \citealt{hendry05}), and about seven times smaller than the luminous
 event like SN 1998A ($25.4\times\,10^{49}$erg~\msun\,$^{-1}$, \citealt{pastorello05}).}. 

 Though we have an approximate value of the pre-SN radius (128 \rsun\,), it can be seen that the progenitor radius of
 \sn\ is smaller than or comparable to that of low-luminosity SNe 1997D ($\sim 300$\rsun\,, \citealt{turatto98}), 2005cs 
 ($100-600$\rsun\,, \citealt{pastorello09}, \citealt{utrobin08}), 1999br ($114$\rsun\,, 
 \citealt{zampieri03}) and much smaller than that of normal Type IIP SNe like 1992H 
 ($\textgreater 600$\rsun\,, \citealt{clocchiatti96}), 2004et ($\sim 530$\rsun\,, \citealt{misra07}). 
 We also note that \sn\ progenitor is only slightly larger than the blue supergiant progenitors of SNe 1987A and 1998A 
 \citep{pastorello05}.

 Information about the metallicity of the SNe location is essential to constrain the triggering mechanism of 
 the SN explosion \citep{heger03}. The oxygen abundance [O/H] of \sn\ is about 8.44 dex \footnote{Derived from the
 [O/H]($= 12 + log(N_O/N_H)$)-$M_{B}$ relation given by \citet{pilyugin04} for a given deprojected radius and galaxy
 type. The host of \sn\ is a spiral galaxy of SBbc Type and the SN location is $\sim 1.8\arcmin$ away from 
 the center of the host. This corresponds to a deprojected distance $\sim 7.03$ kpc.} 
 which is marginally lower than the solar value of 8.65 dex \citep{asplund09}. 
 A comparison \footnote{We have taken the [O/H] value from \citet{smartt09}, though for a few cases the values are estimated using 
 the [O/H]$-M_B$ relation mentioned in \citet{pilyugin04}. The [O/H] for normal Type IIP SNe 2008gz, 1999em and low-luminosity 
 SNe 1999eu, 2005cs is about 8.6. For the low-luminosity SNe 1994N, 1999br, 2001dc, 2003gd and 2004dj, 1997D and the normal 
 SN 2004et, it is about 8.4. In many cases, we have adopted the [O/H] value from \citet{smartt09}. For some cases we have 
 estimated the value by adopting the [O/H]$-M_B$ relation mentioned in \citet{pilyugin04}.}
 of [O/H] of all the events in our sample indicate that there is no clear trend in the type of event and the metallicity.

 The photometric and spectroscopic comparisons of \sn\ with some Type II SNe covering a wide range of physical parameters puts
 observational constraints on the nature of the progenitor of this event, pointing toward a star that was more compact than a
 typical M-Type red supergiants, and closer to a blue supergiant, may be an yellow supergiant. However, the direct detection of 
 the progenitors of a few faint SNe IIP in pre-explosion archive {\it Hubble Space Telescope} images seems to contradict
 this conclusion (see \citealt{smartt09} for a review).
 Our spectroscopic study suggests that in \sn\ the hydrogen envelope ejected by the explosion is smaller than in most
 Type II events but larger than that ejected in low luminosity events (such as SN 1997D, SN 1999br and SN 1999eu).
 This may be due to a significant mass loss of the parent star in the latest stages of its life. This conclusion is also
 partially supported by the upper limit mass loss rate of the progenitor revealed from the X-ray study (\S\ref{res:proge}).
 In view of the upper mass limit of 20 M$_\odot$ progenitor, occurrence of the event in a sub-solar metallicity region and
 following the evolution of a single massive star as a function of metallicity \citep{heger03}, we rule out the possibility
 of ``fall back of ejecta to BH'' scenario in case of \sn\,, supporting the scenario of a weak explosion of a relatively
 compact progenitor.  


\section{conclusion} \label{concl} 

 Low-luminosity Type IIP SNe belong to a poorly known class of events due to the unknown nature of their progenitors
 as well as the explosion mechanism. Spectroscopic as well as photometric characteristics of these events are significantly 
 different from the normal and peculiar Type IIP SNe. The number of such underluminous events discovered so far is relatively small. 
 This could plausibly be a selection effect. If we confine the search of core-collapse SNe to a small volume, the majority of SNe
 discovered in the local universe are of Type IIP ($\sim 48.2\%$, \citealt{smith10}) and probably a large fraction of them would
 turn out to be underluminous. Incidentally SN 1999br, SN 2005cs and SN 2009md are the only events of this group to have been 
 discovered soon after the core-collapse and whose data are publicly available. 

 In this study we have reported the results of an extensive photometric and spectroscopic 
 follow-up campaign for \sn\, in the ultraviolet, optical and  near-infrared domains.
 The SN was also observed (although not detected) at X-ray and radio wavelengths. 
 The SN was observed soon after its explosion, during the fast rise of the light curve to the optical maximum. 
 Evidence for a shock breakout in \sn\ was primarily derived through the analysis of the ROTSE-IIIb  $R$-band  light curve. 
 An upper limit of 16.16 mag in the $R$-band was estimated about two days before the discovery.

 Modeling the $R$-band light curve allowed us to estimate a reliable epoch of the shock breakout, with an uncertainty of 
 about two days. The plateau phase in \sn\ lasts about 98 days which is marginally shorter than in normal Type IIP SNe.
 The evolution of the bolometric light curve indicate that the event is somewhat in between the normal and faint Type IIP 
 event. The luminosity of the nebular phase light curve indicates an ejected $^{56}$Ni mass of $\sim0.015\msun\,$,
 a factor two higher than that derived for low-luminosity IIP SNe. The spectroscopic evolution of \sn\ is similar to 
 those of low-luminosity IIP SNe (1997D, 1999br and 1999eu), indicated particularly by the strong presence of \Baii\ lines, 
 the narrow line widths of \ha\ lines, and the expansion velocities of SN during the photospheric phases ($\sim 3000$ \kms)
 and the nebular phases ($\sim$ 1200 \kms). The ejecta kinematics of \sn\ are consistent with less-energetic
 ($\sim 5\times10^{50}$ erg) Type IIP SNe. 

 Spectroscopically \sn\ appears to be like low-luminosity SNe IIP, but photometrically it appears close to a normal type IIP event.
 
 From the light curve and spectra of \sn\,, we could determine quite accurate values of the observed properties such as
 plateau duration, mid-plateau luminosity as well as the photospheric velocity and this has helped us to comment on 
 the properties of the explosion and the progenitor star. Using semi-analytical formulae by \citet{litvinova85}, we could estimate
 approximate values of the explosion energy $\sim5.4\times10^{50}$ erg, the ejected mass $\sim17\msun$ and the pre-SN 
 radius $\sim127\rsun$. The explosion energy of \sn\ is smaller than the normal ($\ge 10^{51}$ erg) Type IIP events,
 although higher than that estimated in underenergetic ($\sim 10^{50}$ erg) SNe IIP. We could provide an upper limit to the
 mass loss rate of the progenitor as $5\times10^{-6}~M_{\odot}~{\rm yr}^{-1}~(v_{\rm w}/10~{\rm km~s}^{-1})$ whereas the
 upper limit for the main-sequence mass of the progenitor star is estimated as 20 \msun.

\acknowledgments 

 We thank all the observers at Aryabhatta Research Institute of Observational 
 Sciences (ARIES) who provided their valuable time and support for the 
 observations of this event. We are thankful to the observing 
 staffs of ROTSE, REM, 2-m IGO, 3.6-m NTT, 6-m BTA and 9.2-m HET for their kind 
 cooperation in observation of \sn\,.  
 This work was supported by the grant RNP 2.1.1.3483 of the Federal Agency of 
 Education of Russia. Timur A. Fatkhullin and Alexander S. Moskvitin were supported 
 by the grant of the President of the Russian Federation (MK-405.2010.2). This is also 
 partially based on observations collected at the European Southern Observatory, 
 Chile under the program 083.D-0970(A). Stefano Benetti and Filomena Bufano are 
 partially supported by the 
 PRIN-INAF 2009 with the project ``Supernovae Variety and Nucleosynthesis Yields''. 
 The research of J. Craig Wheeler is supported in part by NSF Grant AST-0707669 and by 
 the Texas Advanced Research Program grant ASTRO-ARP-0094. 
 This research is supported by NASA grant NNX08AV63G and NSF grant PHY-0801007. 
 This work is partially based on observations made with the REM Telescope, INAF Chile. 
 This research has made use of data obtained through the High Energy Astrophysics 
 Science Archive Research Center Online Service, provided by the NASA/Goddard 
 Space Flight Center. We are indebted to the Indo-Russian (DST-RFBR) project No. 
 RUSP-836 (RFBR-08-02:91314) for the completion of this research work. 
 

\end{document}

%% file: uvot.tex

\begin{deluxetable*}{c c c c c c c c c}
  \tablecolumns{9}
  \tablewidth{0pt}
  \tablecaption{The {\it swift}/UVOT photometric observations of SN 2008in.\label{tab:uvotsn}}
  \tabletypesize{\scriptsize}
  \tablehead{
    \colhead{UT Date}&
    \colhead{JD}&
    \colhead{Phase\tablenotemark{a}}&
    \colhead{$uvw2$}&
    \colhead{$uvm2$}&
    \colhead{$uvw1$}&
    \colhead{$u$}&
    \colhead{$b$}&
    \colhead{$v$} \\
   (yy/mm/dd)& 2454000+& (day)& (mag)& (mag)& (mag)& (mag)& (mag)& (mag)
   }
  \startdata
  2008/12/30.08 & 830.58 &  5.00 & 14.97$\pm$0.05 & 14.65$\pm$0.04 & 14.45$\pm$0.05 & 14.12$\pm$0.04 & 15.24$\pm$0.04 & 15.20$\pm$0.04 \\
  2008/12/30.60 & 831.11 &  5.53 & 15.13$\pm$0.05 & 14.85$\pm$0.04 & 14.54$\pm$0.04 & 14.16$\pm$0.04 & 15.20$\pm$0.04 & 15.13$\pm$0.05 \\
  2008/12/31.96 & 832.46 &  6.88 & 15.53$\pm$0.05 & 15.28$\pm$0.09 & 14.84$\pm$0.05 & 14.32$\pm$0.04 & 15.25$\pm$0.04 & 15.15$\pm$0.06 \\
  2009/01/01.83 & 833.30 &  7.72 & 15.72$\pm$0.05 & 15.47$\pm$0.05 & 15.01$\pm$0.05 & 14.34$\pm$0.04 & 15.30$\pm$0.04 & 15.17$\pm$0.04 \\
  2009/01/03.54 & 835.05 &  9.47 & 16.23$\pm$0.05 & 16.07$\pm$0.05 & 15.51$\pm$0.05 & 14.52$\pm$0.04 & 15.35$\pm$0.04 & 15.25$\pm$0.04 \\
  2009/01/07.79 & 839.29 & 13.71 & 17.87$\pm$0.08 & 17.88$\pm$0.11 & 16.96$\pm$0.07 & 15.15$\pm$0.04 & 15.49$\pm$0.04 & 15.25$\pm$0.05 \\
  2009/01/11.15 & 842.64 & 17.06 & 18.81$\pm$0.11 & 18.66$\pm$0.14 & 17.80$\pm$0.09 & 15.74$\pm$0.04 & 15.62$\pm$0.04 & 15.27$\pm$0.04 \\
  2009/01/19.79 & 851.30 & 25.72 &       $-$      &       $-$      & 18.67$\pm$0.16 & 16.67$\pm$0.05 & 16.00$\pm$0.04 & 15.36$\pm$0.05 \\
  2009/01/22.00 & 884.53 & 58.95 &       $-$      &       $-$      &       $-$      & 17.92$\pm$0.07 & 16.77$\pm$0.05 & 15.69$\pm$0.04 \\
  2009/01/22.73 & 885.23 & 59.65 &       $-$      &       $-$      &       $-$      & 18.02$\pm$0.09 & 16.75$\pm$0.05 & 15.68$\pm$0.05\\
  \enddata
  \tablenotetext{a}{With reference to the explosion epoch JD 2454825.6}
\end{deluxetable*}

%% file: sec_star.tex

\begin{deluxetable*}{cccccccc}
  \tablecolumns{8}
  \tablewidth{0pt}
  \tabletypesize{\scriptsize}
  \tablecaption{The photometric magnitudes of secondary standard stars in the field of SN 2008in.\label{tab:photstar}}
  \tablehead{ 
    \colhead{Star}& 
    \colhead{$\alpha_{\rm J2000}$}& 
    \colhead{$\delta_{\rm J2000}$}& 
    \colhead{$U$}&
    \colhead{$B$}&
    \colhead{$V$}&
    \colhead{$R$}&
    \colhead{$I$}\\
       ID& (h m s)& (\degr\,\arcmin\,\arcsec)& (mag)& (mag)& (mag)& (mag)& (mag)
     }
    \startdata
      1&12 21 38&04 30 26&16.99$\pm$0.03&17.69$\pm$0.03&17.33$\pm$0.05&16.96$\pm$0.06&16.59$\pm$0.05\\
      2&12 21 40&04 28 18&      $-$     &19.49$\pm$0.05&18.38$\pm$0.04&17.85$\pm$0.01&17.40$\pm$0.02\\
      3&12 21 44&04 25 24&17.91$\pm$0.04&18.09$\pm$0.02&17.65$\pm$0.01&17.30$\pm$0.02&16.94$\pm$0.01\\
      4&12 21 48&04 32 27&18.38$\pm$0.11&17.61$\pm$0.01&16.02$\pm$0.01&15.10$\pm$0.01&14.34$\pm$0.01\\
      5&12 21 50&04 34 20&19.10$\pm$0.21&18.45$\pm$0.04&17.25$\pm$0.01&16.55$\pm$0.01&15.99$\pm$0.01\\
      6&12 21 57&04 34 38&16.06$\pm$0.01&15.72$\pm$0.00&14.88$\pm$0.00&14.36$\pm$0.01&13.94$\pm$0.00\\
      7&12 22 04&04 32 30&16.66$\pm$0.02&16.30$\pm$0.00&15.48$\pm$0.01&15.00$\pm$0.01&14.61$\pm$0.01\\
      8&12 22 05&04 30 51&18.05$\pm$0.06&18.17$\pm$0.01&17.60$\pm$0.04&17.25$\pm$0.03&16.93$\pm$0.01\\
      9&12 21 43&04 31 54&19.56$\pm$0.41&19.20$\pm$0.05&17.55$\pm$0.01&16.53$\pm$0.01&15.51$\pm$0.01\\
     10&12 21 33&04 29 29&19.11$\pm$0.43&18.96$\pm$0.02&18.55$\pm$0.02&18.21$\pm$0.05&17.89$\pm$0.02\\
  \enddata

  \tablecomments{Errors in magnitude represent RMS scatter of the night-to-night repeatability  
           over entire period of SN monitoring.}
\end{deluxetable*}

%% file: phot_sn.tex

\begin{deluxetable*}{c c c c c c c c c }
  \tablecolumns{9}
  \tablewidth{0pt}
  \tablecaption{The $UBVRI$ photometry of SN 2008in.\label{tab:photsn}}
  \tabletypesize{\scriptsize}
 \tablehead{
  \colhead{UT Date}&
  \colhead{JD}&
  \colhead{Phase\tablenotemark{a}}&
  \colhead{$U$}&
  \colhead{$B$}&
  \colhead{$V$}&
  \colhead{$R$}&
  \colhead{$I$}&
  \colhead{Seeing\tablenotemark{c}} \\
  (yy/mm/dd)& 2454000+& (day)& (mag)& (mag)& (mag)& (mag)& (mag)&(\arcsec)
  }
  \startdata
   2008/12/30.00 & 830.50 & $+$5 & 14.23$\pm$0.01 & 15.33$\pm$0.01 & 15.07$\pm$0.01 & 14.86$\pm$0.01 & 14.86$\pm$0.01 & 2.9\\
   2009/01/01.98 & 833.48 & $+$8 & 14.39$\pm$0.01 & 15.54$\pm$0.01 & 15.17$\pm$0.01 & 14.65$\pm$0.01 & 14.98$\pm$0.01 & 1.3\\
   2009/01/09.02 & 840.52 &$+$15 & 15.47$\pm$0.02 & 15.76$\pm$0.02 & 15.21$\pm$0.02 & 14.89$\pm$0.02 & 14.87$\pm$0.03 & 2.5\\
   2009/01/23.95 & 855.45 &$+$30 & 16.85$\pm$0.04 & 16.36$\pm$0.01 & 15.38$\pm$0.02 & 15.00$\pm$0.01 & 14.85$\pm$0.01 & 2.1\\
   2009/01/26.02 & 857.52 &$+$32 & 16.90$\pm$0.04 & 16.41$\pm$0.02 & 15.39$\pm$0.01 & 14.98$\pm$0.01 & 14.85$\pm$0.02 & 3.1\\
   2009/01/27.01 & 858.33 &$+$33 & 16.81$\pm$0.02 & 16.41$\pm$0.01 & 15.39$\pm$0.01 & 15.00$\pm$0.01 & 14.83$\pm$0.01 & 2.8\\
   2009/01/28.83 & 860.33 &$+$35 & 16.83$\pm$0.03 & 16.48$\pm$0.01 & 15.43$\pm$0.01 & 15.04$\pm$0.01 & 14.89$\pm$0.01 & 2.2\\
   2009/01/30.88 & 862.38 &$+$37 & 17.16$\pm$0.04 & 16.53$\pm$0.01 & 15.43$\pm$0.01 & 15.02$\pm$0.01 & 14.85$\pm$0.01 & 2.4\\
   2009/02/02.89 & 864.39 &$+$39 & 17.11$\pm$0.04 & 16.53$\pm$0.01 & 15.42$\pm$0.01 & 14.99$\pm$0.01 & 14.78$\pm$0.01 & 2.0\\
   2009/02/15.82 & 878.32 &$+$53 & 17.58$\pm$0.05 & 16.87$\pm$0.01 & 15.43$\pm$0.02 & 14.76$\pm$0.01 & 14.82$\pm$0.01 & 2.8\\
   2009/02/16.96 & 879.46 &$+$54 & 17.74$\pm$0.06 & 16.84$\pm$0.01 & 15.54$\pm$0.01 & 15.09$\pm$0.02 & 14.83$\pm$0.01 & 3.1\\
   2009/02/17.79 & 880.29 &$+$55 & 17.79$\pm$0.05 & 16.87$\pm$0.01 & 15.55$\pm$0.01 & 15.10$\pm$0.02 & 14.85$\pm$0.01 & 3.1\\
   2009/02/20.87 & 883.37 &$+$58 & 17.92$\pm$0.07 & 16.94$\pm$0.01 & 15.59$\pm$0.02 & 15.09$\pm$0.01 & 14.84$\pm$0.01 & 3.1\\
   2009/02/23.77 & 886.27 &$+$61 & 18.05$\pm$0.04 & 17.08$\pm$0.01 &        $-$     & 15.13$\pm$0.01 & 14.89$\pm$0.01 & 3.1\\
   2009/02/25.81 & 888.31 &$+$63 & 18.13$\pm$0.05 & 17.05$\pm$0.01 & 15.60$\pm$0.01 & 15.14$\pm$0.01 & 14.92$\pm$0.01 & 2.9\\
   2009/02/26.80 & 889.30 &$+$64 &        $-$     &        $-$     & 15.62$\pm$0.01 & 15.15$\pm$0.01 &        $-$     & 3.8\\
   2009/03/02.78 & 893.28 &$+$68 &        $-$     & 17.08$\pm$0.01 & 15.63$\pm$0.01 & 15.17$\pm$0.01 & 14.90$\pm$0.00 & 4.4\\
   2009/03/04.77 & 895.27 &$+$70 &        $-$     & 17.15$\pm$0.01 & 15.66$\pm$0.01 & 15.18$\pm$0.01 & 14.94$\pm$0.01 & 2.4\\
   2009/03/09.76 & 900.27 &$+$75 & 18.72$\pm$0.14 & 17.31$\pm$0.02 & 15.68$\pm$0.01 &        $-$     & 15.08$\pm$0.01 & 2.3\\
   2009/03/15.27 & 905.77 &$+$80 &        $-$     & 17.20$\pm$0.02 &        $-$     & 15.21$\pm$0.01 & 14.92$\pm$0.01 & 2.2\\
   2009/03/24.80 & 915.30 &$+$90 & 18.73$\pm$0.08 & 17.42$\pm$0.01 & 15.94$\pm$0.01 & 15.48$\pm$0.01 & 15.15$\pm$0.01 & 2.5\\
   2009/04/03.75 & 925.25 &$+$100& 19.28$\pm$0.15 & 17.66$\pm$0.02 & 16.10$\pm$0.01 & 15.57$\pm$0.01 & 15.31$\pm$0.01 & 2.2\\
   2009/04/05.61 & 927.22 &$+$102&        $-$     &        $-$     &        $-$     & 15.67$\pm$0.01 & 15.41$\pm$0.01 & 2.5\\
   2009/04/10.67 & 932.33 &$+$107&        $-$     & 18.38$\pm$0.06 & 16.76$\pm$0.01 & 16.20$\pm$0.01 & 15.84$\pm$0.02 & 2.7\\
   2009/04/17.60 & 939.25 &$+$114&        $-$     & 19.83$\pm$0.06 & 17.91$\pm$0.01 & 17.13$\pm$0.01 & 16.68$\pm$0.01 & 2.8\\
   2009/04/22.68 & 944.11 &$+$118&        $-$     & 20.30$\pm$0.16 & 18.10$\pm$0.03 & 17.21$\pm$0.03 & 16.85$\pm$0.04 & 3.5\\
   2009/04/28.71 & 950.17 &$+$125&        $-$     & 20.40$\pm$0.09 &        $-$     & 17.24$\pm$0.01 & 16.74$\pm$0.01 & 2.2\\
   2009/05/01.72 & 953.10 &$+$127&        $-$     & 20.22$\pm$0.11 & 18.06$\pm$0.04 & 17.27$\pm$0.04 & 16.70$\pm$0.04 & 2.7\\
   2009/05/23.69 & 975.18 &$+$150&        $-$     &        $-$     & 18.30$\pm$0.04 & 17.47$\pm$0.02 & 17.00$\pm$0.03 & 2.7\\
   2009/05/25.70 & 977.21 &$+$152&        $-$     & 20.21$\pm$0.12 & 18.35$\pm$0.02 & 17.45$\pm$0.01 &        $-$     & 2.6\\
   2009/05/26.71 & 978.22 &$+$153&        $-$     & 20.15$\pm$0.17 & 18.33$\pm$0.04 & 17.49$\pm$0.02 & 17.06$\pm$0.03 & 2.8\\
   2009/06/14.69 & 997.19 &$+$172&        $-$     &        $-$     & 18.39$\pm$0.04 &        $-$     &        $-$     & 2.9\\
   2010/01/17.98 &1214.49 &$+$389&        $-$     &        $-$     &        $-$     & 19.69$\pm$0.07 & 19.27$\pm$0.13 & 1.9\\
   2010/01/19.91 &1216.42 &$+$391&        $-$     & 21.47$\pm$0.12 & 20.34$\pm$0.05 & 19.76$\pm$0.05 & 19.52$\pm$0.07 & 2.3\\
   2010/02/13.91 &1241.42 &$+$416&        $-$     &        $-$     & 20.57$\pm$0.07 & 19.77$\pm$0.06 &        $-$     & 2.2\\
  \enddata                                                                                                  
  \tablenotetext{a}{With reference to the explosion epoch JD 2454825.6.}
  \tablenotetext{c}{FWHM of the stellar PSF at $V$ band. The nights for which we do not have any
  observation in $V$ band, PSF is found through interpolation.}

  \tablecomments{The photometric observations are taken with the 1-m Sampurnanand Telescope, ARIES, Nainital. 
    Errors in magnitude denote $1\sigma$ uncertainty.}
\end{deluxetable*}

%% file: rem.tex

\begin{deluxetable*}{c c c c c}
  \tablecolumns{5}
  \tablewidth{0pt}
  \tablecaption{The REM near-infrared observation of SN 2008in.\label{tab:remsn}}
  \tabletypesize{\scriptsize}
  \tablehead{
     \colhead{UT Date}&
     \colhead{JD}&
     \colhead{Phase\tablenotemark{a}}&
     \colhead{$J$}&
     \colhead{$H$}\\
     (yy/mm/dd)& 2454000+& (day)& (mag)& (mag)
    }
  \startdata
  2008/12/29.85 & 830.36 &   4.78 & 14.66$\pm$0.054 & 14.55$\pm$0.064 \\
  2009/01/03.87 & 835.37 &   9.78 & 14.60$\pm$0.077 & 14.38$\pm$0.058 \\
  2009/01/09.73 & 841.23 &  15.65 & 14.39$\pm$0.059 & 14.16$\pm$0.072 \\
  2009/01/14.73 & 846.23 &  20.65 & 14.47$\pm$0.107 & 14.32$\pm$0.150 \\
  2009/01/19.78 & 851.24 &  25.66 & 14.54$\pm$0.150 &       $-$       \\
  2009/01/24.76 & 856.26 &  30.67 & 14.25$\pm$0.077 & 14.11$\pm$0.087 \\
  2009/01/29.87 & 861.37 &  35.79 & 14.37$\pm$0.050 & 14.14$\pm$0.065 \\
  2009/02/05.66 & 868.16 &  42.58 & 14.43$\pm$0.067 & 14.24$\pm$0.107 \\
  2009/02/11.69 & 874.19 &  48.61 & 14.20$\pm$0.107 & 14.59$\pm$0.241 \\
  2009/02/16.72 & 879.22 &  53.64 & 14.43$\pm$0.050 & 14.35$\pm$0.097 \\
  2009/02/21.75 & 884.25 &  58.67 &       $-$       & 14.18$\pm$0.063 \\
  2009/02/27.75 & 890.25 &  64.67 &       $-$       & 14.44$\pm$0.077 \\
  2009/03/04.89 & 895.40 &  69.82 & 14.40$\pm$0.074 & 14.12$\pm$0.085 \\
  2009/03/10.74 & 901.24 &  75.66 &       $-$       & 14.36$\pm$0.190 \\
  2009/03/21.71 & 912.21 &  86.63 &       $-$       & 14.50$\pm$0.090 \\
  2009/03/26.73 & 917.23 &  91.65 &       $-$       & 14.41$\pm$0.073 \\
  2009/04/05.79 & 927.29 & 101.71 &       $-$       & 15.00$\pm$0.147 \\
  2009/04/10.80 & 932.30 & 106.71 &       $-$       & 14.94$\pm$0.171 \\
  2009/04/15.80 & 937.30 & 111.72 &       $-$       & 14.86$\pm$0.173 \\
  2009/04/20.52 & 942.02 & 116.44 &       $-$       & 15.23$\pm$0.232 \\
  \enddata
  \tablenotetext{a}{With reference to the explosion epoch JD 2454825.6.}
\end{deluxetable*}

%% file: speclog.tex

\begin{deluxetable*}{lc cc cc cc cc}
  \tablecolumns{10}
  \tablewidth{0pt}
  \tablecaption{Journal of spectroscopic observations of SN 2008in.\label{tab:speclog}}
  \tabletypesize{\scriptsize}
  \tablehead{
    \colhead{UT Date}& 
    \colhead{JD}&
    \colhead{Phase\tablenotemark{a}}&
    \colhead{Range}&
    \colhead{Telescope\tablenotemark{b}}&
    \colhead{ Grating}&
    \colhead{ Slit width}&
    \colhead{ Dispersion}&
    \colhead{ Exposure}&
    \colhead{S/N\tablenotemark{c}} \\
    (yy/mm/dd/hh.hh)& 2454000+& (days)& \mum&     & (gr mm$^{-1}$)&  (\arcsec)& (\AA\,pix$^{-1}$)& (s)&(pix$^{-1}$)
  }
  \startdata

        2008/12/31/22.80& 832.45& +7 & 0.45$-$1.00& HET& 300& 1.0 & 5.0& 600&45\\
        2009/01/07/22.31& 839.43& +14 & 0.45$-$1.00& HET& 300& 1.0 & 5.0& 600&43\\
    
        2009/02/17/19.74& 880.32& +54 & 0.38$-$0.68& IGO& 600& 1.5 & 1.4& 1800&34\\
    
        2009/02/22/22.50& 885.44& +60 & 0.45$-$1.00& HET& 300& 1.0 & 5.0& 600&35\\
    
        2009/03/22/20.45& 913.35& +87 & 0.38$-$0.68& IGO& 600& 1.5 & 1.4& 1800&38\\
        2009/03/25/22.13& 916.42& +90 & 0.38$-$0.68& IGO& 600& 1.5 & 1.4& 2700&35\\
        2009/04/03/19.14& 925.26& +99 & 0.38$-$0.68& IGO& 600& 1.5 & 1.4& 1800&35\\

        2009/04/03/20.98& 925.33& +99 & 0.60$-$1.00& BTA& 550& 2.1 & 3.5& 3$\times$900&60\\

        2009/04/22/15.49& 944.14& +118& 0.38$-$0.68& IGO& 600& 1.5 & 1.4& 1800&11\\
    
        2009/04/23/18.00& 945.25& +119& 0.38$-$0.79& BTA& 600& 2.1 & 3.5& 3$\times$900&65\\
    
        2009/05/17/02.38& 968.10& +143& 0.33$-$0.80& NTT& 300& 1.0 & 4.0& 2700&59\\
  \enddata

  \tablenotetext{a}{With reference to the explosion epoch JD 2454825.6.}
  \tablenotetext{b}{HET : LRS on 9.2-m Hobby Eberly Telescope; IGO : IFOSC on
  2-m IUCAA Girawali Observatory, India; BTA : SCORPIO on 6-m Big Telescope Alt-azimuth,
  Special Astrophysical Observatory, Russia; 
  NTT : EFOSC2 on 3.6-m New Technology Telescope, ESO, Chile.}
  \tablenotetext{c}{At 0.6 \mum.}
\end{deluxetable*}